\documentclass[showpacs,amsmath,amssymb,prl,superscriptaddress,footinbib,twocolumn]{revtex4-1}
\usepackage{amssymb}
\usepackage{graphics}
\usepackage{epstopdf}
\usepackage{epsfig}
\usepackage{dcolumn}
\usepackage{bm}
\renewcommand{\vec}[1]{\mathbf{#1}}
 
\newcommand{\abs}[1]{\left| #1 \right|} 
 
\let\baraccent=\= 
\renewcommand{\=}[1]{\stackrel{#1}{=}} 
 
\newcommand{\appropto}{\mathrel{\vcenter{ \offinterlineskip\halign{\hfil$##$\cr\propto\cr\noalign{\kern2pt}\sim\cr\noalign{\kern-2pt}}}}}

\usepackage{color}
 \definecolor{blue}{rgb}{0,0,1} 
 \definecolor{sepia}{rgb}{0,0.8,0.2}
 \definecolor{redi}{rgb}{0.8176,0.0078,0.0078}

\usepackage{ulem}
\begin{document}

\title{Electrostatic bending response of a charged helix }



\author{A. V.  Zampetaki}
\author{J. Stockhofe}
\affiliation{Zentrum f\"{u}r Optische Quantentechnologien, Universit\"{a}t Hamburg, Luruper Chaussee 149, 22761 Hamburg, Germany}
\author{P. Schmelcher}
\affiliation{Zentrum f\"{u}r Optische Quantentechnologien, Universit\"{a}t Hamburg, Luruper Chaussee 149, 22761 Hamburg, Germany}
\affiliation{The Hamburg Centre for Ultrafast Imaging, Luruper Chaussee 149, 22761 Hamburg, Germany}

\date{\today}

\begin{abstract}
We explore the electrostatic bending response of a chain of charged particles confined on
a finite helical filament. We analyze how the energy difference $\Delta E$ between
the bent and the unbent helical chain scales with the length of the helical segment and the radius of curvature 
and identify features that are not captured by the standard notion of the bending rigidity, normally
used as a measure of bending tendency in the linear response regime. Using instead $\Delta E$ to characterize the bending response of the helical 
chain we identify two regimes with qualitatively different bending behaviors for the ground state configuration: the regime of small
and the regime of large radius-to-pitch ratio, respectively. Within the former regime, $\Delta E$ changes smoothly with the variation of the system parameters.
Of particular interest are its oscillations with the number of charged particles encountered for commensurate fillings which yield length-dependent oscillations 
in the preferred bending direction of the helical chain. We show that the origin of these oscillations is the non-uniformity of the 
charge distribution caused by the long-range character of the Coulomb interactions and the finite length of the helix. In the second regime of large values of the 
radius-to-pitch ratio, sudden changes in the ground state structure of the charges occur as the system parameters vary, 
leading to complex and discontinuous variations in the ground state bending response $\Delta E$.

\end{abstract}

\maketitle

\begin{center}
 \textbf{I. INTRODUCTION} 
\end{center}
The atomic structure of biological macromolecules often involves a single or multiple interwoven helical chains. The $\alpha$-helix of several polypeptides and polyaminoacids,
the double stranded helix of the DNA and the triple helix of the  collagen are some of the most prominent examples \cite{Kornyshev2007}. These biological helices
are rarely found in a straight conformation. Usually they appear to be bent or even supercoiled, rendering the investigation of their bending response an important 
subject of research \cite{Kornyshev2007, Marco1994, Manning1996, Benham2005, Balaeff2006}. The majority of these studies \cite{Marco1994, Manning1996, Benham2005, Balaeff2006} 
handle the macromolecules as elastic rods within the so-called continuum Kirchhoff theory \cite{Antman2005} and ignoring
further interactions between the particles (atoms, ions or molecules) constituting the helical chain.

On the other hand many macromolecules such as the DNA, as well as several synthetic polyelectrolytes and proteins are strongly charged (especially when dissolved)
making it essential to account for the electrostatic interactions between the particles constituting them.
Such interactions have been mostly treated within models of infinite, homogeneously charged membranes or
cylinders (polyelectrolytes) \cite{Hagerman1983, Winterhalter1988, Nguyen1999,Lau1998, Han2003, Manghi2004, Dobrynin2006, Trizac2016} 
and very rarely in terms of  helical structures \cite{Tan2008,Sivara2008} in an electrolyte solution. 
Among the most striking results
is the prediction of a negative electrostatic contribution to the bending rigidity for uniformly charged membranes and polyelectrolytes \cite{Nguyen1999}.  This takes place 
for sufficiently thick membranes (or cylinders) on whose surface multivalent counterions crystallize into so-called Wigner crystals \cite{Wigner1932}.

A negative contribution to the bending stiffness is also predicted for the case of dipolar interactions in magnetosome chains modeling certain bacteria \cite{Hanzlik1997,Kiani2015}.
When such systems are treated as  chains of discrete magnetized particles it is found that for  high enough particle numbers a bending of the chain is highly preferable, 
leading even to the formation of closed rings \cite{Kiani2015,Messina2014,Wei2011}.

Returning to the case of charged helical structures it has been shown that, even in the absence of an electrolyte solution, charges confined on a helical trap can display
interesting properties \cite{Kibis1992,Schmelcher2011,Plettenberg2016,Zampetaki2013,Zampetaki2015a,Zampetaki2015b, Zampetaki2017} such as 
effective oscillatory interactions \cite{Kibis1992,Schmelcher2011}, band structure inversions \cite{Zampetaki2015a} and unconventional pinned-to-sliding transitions \cite{Zampetaki2017}. 
The helical confining manifold essential for the emergence of such features can have different shapes.
An infinite homogeneous \cite{Kibis1992,Schmelcher2011,Plettenberg2016} or inhomogeneous helix \cite{Zampetaki2013} as well as a closed helix \cite{Zampetaki2015a,Zampetaki2015b, Zampetaki2017} have all been used 
as prerequisites in order to highlight different aspects of the helical confinement. 
The question, however, of what is the energetically preferable conformation of the helical filament given the Coulomb interactions between the charges confined on it still remains
open. We attempt here to address some aspects of this question.

In particular, we study in this work the bending response of a system of identical charges confined on a finite helical segment. Such a bending response
is usually described in terms of the bending rigidity (bending stiffness) and the persistence length 
\cite{Marco1994,Antman2005, Hagerman1983, Winterhalter1988,Lau1998,Nguyen1999, Han2003,Manghi2004,Dobrynin2006,Trizac2016,Sivara2008,Kiani2015}. 
The definition of these quantities relies on a simple scaling of the energy difference $\Delta E$ between
the bent and the unbent configuration with the system length $L$ and the radius of curvature $R$ \cite{Nguyen1999,Kiani2015}. 
We show that for our system $\Delta E$ exhibits an unconventional scaling with $R$, $L$ such that its bending response cannot be captured by the standard definition of the bending rigidity.
In our analysis we therefore focus on the bending-induced energy difference $\Delta E$.

In contrast
to other studies \cite{Hagerman1983, Winterhalter1988, Nguyen1999,Han2003} we do not assume a uniform distribution of charged particles on the helix segment but use their actual, numerically obtained ground state (GS) which in view 
of the system's finite length (fixed boundary conditions) is in general non-uniform. For small values of the radius-to-pitch ratio this GS can be relatively easily identified since it 
constitutes the only equilibrium state of the system. In this regime we show that the bending-induced energy difference $\Delta E$ of the helical chain varies smoothly with the system parameters,
taking values of varying signs and magnitudes. Remarkably, we find that for intermediate values of the radius-to-pitch ratio and in the special case of commensurate fillings 
 $\Delta E$ oscillates with the system size. These oscillations can be perceived as oscillations of the preferred bending direction,
reminiscent of existing results about the bistability of helical filaments \cite{Goldstein2000}.  We identify and explain the origin of such oscillations employing both analytical 
and numerical tools. It turns out that they originate from the interplay between the long-range character of interactions and the fixed boundaries, causing a non-uniform 
distribution of charges on the helical segment.

For larger values of the radius-to-pitch ratio a plethora of equilibrium states arises leading to energy crossings signifying 
discontinuous changes of the GS structure. Consequently, the bending-induced energy shift of the GS $\Delta E$ also varies discontinuously with the system parameters.

This paper is structured as follows. We begin in Sec. II by introducing our system of a finite helical chain of charges.
We continue in Sec. III with some analytical estimations of the relevant energy difference $\Delta E$, discussing also its scaling with $R,~L$.
Section IV contains the major numerical results of this work regarding the behaviour of $\Delta E$, quantifying the bending response for different but overall not too large values of the radius-to-pitch ratio, different fillings and different system sizes. 
In the subsequent sections V,VI and VII we provide a number of extensions of these results,
studying in particular the influence of the interaction range (Sec. V) and going to larger values of the radius-to-pitch ratio (Sec. VI) and beyond the weakly-bent response regime
(Sec. VII).
Finally, in Sec. VIII we present our conclusions and give an outlook for future work. 

\begin{figure}[htbp]
\begin{center}
\includegraphics[width=0.8\columnwidth]{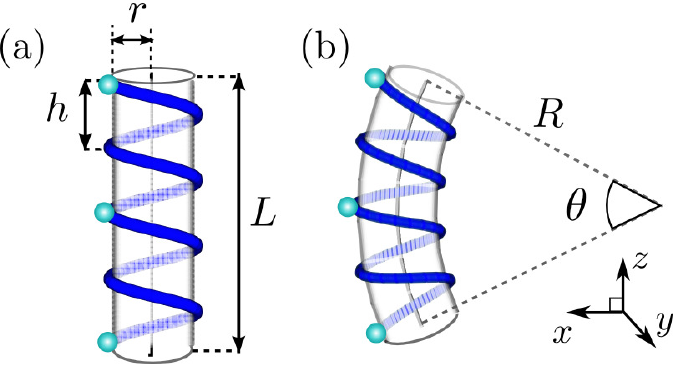}
\end{center}
\caption{\label{syst1} (color online)  Schematic illustration  of the bending of a helix with parameters $r,h,L$ commensurately filled ($\nu=(N-1)/M=1/2$) with charged particles.
For better visualization the final curvature (controlled by the radius of curvature $R$) and therefore also
the degree of bending characterized by the angle $\theta$ have been exaggerated. Note that the helix is assumed here to bend  towards the positive 
$x$-direction.}
\end{figure}

 \begin{center}
 { \textbf{II. SETUP, CONFIGURATIONS AND ENERGETICS OF THE HELICAL CHAIN}}
\end{center}
We consider a system of $N$ identical charged particles of mass $m$  each, interacting through repulsive Coulomb interactions  and confined to move 
on a 1D finite helical segment of length $L$ (Fig.\ref{syst1} (a)) parametrized as

\begin{equation}
 \vec{r} (u)= \begin{pmatrix}
 r \cos u \\
 r \sin u\\
\frac{h}{2\pi} u
\end{pmatrix}, \quad u \in \left[0, \Phi\right]
\label{te1}
\end{equation}
with $r$ denoting the radius of the helix, $h$ being its pitch and $\Phi=2 \pi M$ referring to the angle length of the helix, 
proportional to the number of windings $M$. Note that the length $L$ is connected to $\Phi$ through the pitch, i.e. $L=\frac{h}{2\pi}\Phi$.

The confinement in the 1D helical manifold results in an oscillatory effective interaction potential \cite{Schmelcher2011}
\begin{equation}
 V(\{u_i\})=\sum_{\substack{i,j=1 \\ i\neq j}}^N  \frac{\lambda/2}{\sqrt{2r^2\left[1-\cos(u_i-u_j)\right]+\left(\frac{h(u_i-u_j)}{2\pi}\right)^2}},\label{pothe1}
\end{equation}
with $\lambda$ being the coupling constant between the interacting particles localized at angles $u_j$, $j=1,\dots,N$. Due to its oscillatory behaviour this interaction potential can exhibit
multiple wells, supporting classical bound states, depending on the geometry parameters of the helix \cite{Kibis1992, Schmelcher2011}
and in particular on the ratio $f=\frac{2 \pi r}{h}$. 

If $f=0$ the helical segment reduces to a simple line segment of length $L$ whereas in the opposite case $f \rightarrow \infty$ it reduces to a circle of radius $r$. In both of these
limiting cases the interplay of the boundary constraints (hard wall or periodic boundary conditions) with the purely repulsive Coulomb interaction leads to the existence of a single
equilibrium state of the ions (the ground state (GS)) to which the system crystallizes at zero temperature. For values of $f$ of order unity and beyond, the helical geometry 
is imprinted in the interaction potential, leading to the emergence of multiple potential wells which, combined with the hard wall boundary conditions studied here,
yield a plethora of equilibrium states. Notably the number of such states is expected to increase very rapidly both with the number of particles $N$ and with the ratio $f$ 
\cite{Schmelcher2011, Zampetaki2015a, Zampetaki2017}, resulting in a high degree of complexity and making it particularly difficult to identify the GS of the system.

Taking into account the exceptional impact of the helical confinement on the behavior of a chain of charges, 
as well as the ubiquitous emergence of the helical structure in DNA molecules and proteins,
we investigate in the following the bending response of the helical chain of charges for various geometrical parameters and fillings $\nu=(N-1)/M$ \cite{Note1}.
The bending response, i.e. the way in which a system reacts to an infinitesimal bending (Fig. \ref{syst1}(b)) can be quantified by the energy difference 
$\Delta E=E_b-E_a$  between the GS energies of the system after bending, $E_b$, and before  bending, $E_a$.

A negative value of $\Delta E$ is indicative of the system favoring (energetically) a bent conformation 
whereas a positive value stands for the system resisting to bending. For a linear chain of charges, i.e. in the limiting case $f=0$, the energy difference $\Delta E$
is strictly positive since a bending of a linear chain always results in smaller Euclidean interparticle  distances,
yielding therefore an increment of the potential energy $E_b>E_a$. In  contrast, for a finite value of $f$ the helical segment possesses a finite radius $r$ and therefore 
an inner side and an outer side which respond differently to bending. On the inner side, the Euclidean interparticle distances always decrease
(even more than in the case of the line segment) while on the outer side especially for large values of $f$ the Euclidean interparticle distances can increase 
(Fig. \ref{syst1}(b)). Thus, a negative $\Delta E$ (favor of bending) cannot be precluded, with its actual value expected to depend on the imbalance
between the particles occupying the outer and the inner parts of the helix, similarly to the case of charged membranes \cite{Nguyen1999}.

Note that in this work we are interested only in the bending response resulting from the electrostatics of the charges and we do not account for any contribution
relating to elastic properties of the helical filament.

Having obtained the relevant information about our system setup let us now proceed to the presentation and discussion of our results for its
bending response in specific cases, starting with some analytical estimations.

 \begin{center}
 { \textbf{III. ANALYTICAL ESTIMATIONS OF THE BENDING RESPONSE}}
\end{center}

As shown in  Fig. \ref{syst1} the bent helical segment (b) can be perceived as a finite segment of a toroidal helix \cite{Zampetaki2015a, Zampetaki2015b, Zampetaki2017} with a large major radius of curvature $R$.
Such a segment is parametrized as  
\begin{equation}
 \vec{r}_T (u)= \begin{pmatrix}
 \left( R+r \cos u\right)\cos\left(au-a\frac{\Phi}{2}\right)-R \\
 r\sin u\\
 \left( R+r \cos u\right)\sin\left(au-a\frac{\Phi}{2}\right)+\frac{h\Phi}{4\pi}
\end{pmatrix},  u  \in \left[0, \Phi\right].
\label{te2}
\end{equation}
with $R$ denoting the major radius of the torus, 
 $r$ being the radius of the helix (minor radius of the torus), $a=\frac{h}{2\pi R}$ relating to the pitch $h$ of the straight helical segment and $\Phi$ denoting 
 the total angular length, cf. Eq. (\ref{te1}). Within this parametrization $\vec{r}_T (u) \rightarrow \vec{r} (u)$,
 i.e. the bent helical segment reduces to the straight helical segment (Eq.~(\ref{te1})),
 for an infinite radius of curvature $R\rightarrow \infty,a\rightarrow 0$ and $aR=\frac{h}{2\pi}$.
 
 For an infinitesimal bending $R$ should be arbitrarily large ($R \rightarrow \infty$). However, for the sake of our numerical study we consider here large, but finite values. 
 In particular, we choose
 \begin{equation}
 R=\frac{h}{4\pi^2}\bar{R}\Phi, \label{Rcur}
 \end{equation}
 where $\Phi$ is the angular length of the
 segment defined  above  and $\bar{R}$ a large constant ($\bar{R} \gg 1 $). 
 This choice of $R$ allows us to achieve   a specific
 angle of bending $\theta=\frac{L}{R}=\frac{2\pi}{\bar{R}}$ independent of the system  size (see Fig. \ref{syst1} (b)).
 Note that with this definition $\bar{R}=1$ would correspond to
 the largest possible degree of bending, i.e. closing the helical segment into a ring ($\theta=2\pi$).
 
Given the GS configuration $\{\bar{u}_k^{(0)}\}$ of the bent helix $\vec{r}_T$ and using the notation 
 $\bar{u}_{ij}^{(0)}=\bar{u}_i^{(0)}-\bar{u}_j^{(0)}$,
 the square of the Euclidean distance $d_{ij}^2=\abs{\vec{r}_T(\bar{u}_i^{(0)})-\vec{r}_T(\bar{u}_j^{(0)})}^2$ between the particles $i,j$ confined on the toroidal helical segment reads
  \begin{eqnarray}
\frac{d_{ij}^2}{R^2}&=& \left(2+\frac{2r}{R}\cos \bar{u}_i^{(0)}\right)\left(1+\frac{r}{R}\cos \bar{u}_j^{(0)}\right)\left[1-\cos a \bar{u}_{ij}^{(0)}\right] \nonumber\\
&+&2\left(\frac{r}{R}\right)^2\left[1-\cos \bar{u}_{ij}^{(0)}\right]. \label{dis1}
\end{eqnarray}

For the case of very weak bending the radius of curvature $R$ is very large and in particular 
$R \gg h,r$. In such a case the GS configuration of the bent helix  $\{\bar{u}_k^{(0)}\}$ is approximately the same as the GS configuration $\{{u}_k^{(0)}\}$
of a straight helix. Making the simplifying approximation that these GSs are exactly the same, i.e. $\bar{u}_k^{(0)}=u_k^{(0)}~~\forall k$, 
an approximation that as it will be shown in the next sections still captures qualitatively very well the effects encountered in our numerical simulations, 
we  perform a Taylor expansion of Eq.~(\ref{dis1}) both in terms of $\frac{r}{R}$ and in terms of $a=\frac{h}{2\pi R}$.
The latter expansion is eligible
only if the product $a u_{ij}^{(0)} \ll 1$, a condition whose validity, as we will discuss below, depends on the system size. For the moment let us assume 
that it holds, resulting in
 \begin{equation}
1-\cos \frac{h}{2\pi R} u_{ij}^{(0)} \approx\frac{h^2}{8\pi^2R^2}(u_{ij}^{(0)})^2-\frac{h^4}{32\pi^4R^4}(u_{ij}^{(0)})^4. \label{cosap1}
\end{equation}
 
Given this, as well as  $\frac{r}{R} \ll 1$, it turns out that the first non-zero contribution to $D_{ij}=\frac{d_{ij}^2}{R^2}$ is of second order in both $\frac{h}{R},\frac{r}{R}$ and reads
\begin{equation}
D_{ij}^{(2)}=2\left(\frac{r}{R}\right)^2\left[1-\cos(u_{ij}^{(0)})\right]+\left(\frac{h}{2\pi R}\right)^2(u_{ij}^{(0)})^2,
\end{equation}
which is equal to $\abs{\vec{r}(u_i^{(0)})-\vec{r}(u_j^{(0)})}^2/R^2$, i.e. the square of the distance between the same particles when confined on the straight helical segment of Eq.~(\ref{te1}).
The two successive higher order contributions $D_{ij}^{(3)},D_{ij}^{(4)}$ incorporate the effect of the segment's curvature, i.e. the effect of bending and read
\begin{equation}
D_{ij}^{(3)}=\frac{h^2r}{4\pi^2 R^3}\left(\cos u_i^{(0)}+\cos u_j^{(0)}\right)(u_{ij}^{(0)})^2 \label{d3eq1}
\end{equation}
and
\begin{eqnarray}
D_{ij}^{(4)}&=&-\frac{1}{12}\left(\frac{h}{2\pi R}\right)^4(u_{ij}^{(0)})^4 \nonumber \\
&+& \frac{h^2r^2}{4\pi^2 R^4}\cos u_i^{(0)}\cos u_j^{(0)}(u_{ij}^{(0)})^2.\label{d4eq1}
\end{eqnarray}

Let us now have a closer look at the terms constituting  $D_{ij}^{(3)}$ and $D_{ij}^{(4)}$. Obviously these are combinations of $\epsilon_r=\frac{r}{R}$ and $\epsilon_h=\frac{h}{2\pi R}|u_{ij}^{(0)}|$. While the first is always small, the latter depends on the GS configuration $\{u_{i}^{(0)}\}$ as well as on the values of the indices $i,j$ and lies in the interval 
 $\left[0,\frac{h \Phi}{2\pi R}\right]$. For filling $\nu=(N-1)/M$ and an approximately equidistant GS configuration $\{u_{i}^{(0)}=2\pi(i-1)/\nu\}$
 we see that for
 $\abs{i-j}=1$ (nearest neighbors) $\frac{\epsilon_r}{\epsilon_h}\approx \frac{r \nu}{h}$, whereas
 $\abs{i-j}=N-1$ yields $\frac{\epsilon_r}{\epsilon_h}\approx \frac{r}{hM}$. Thus for $\frac{r}{h}\approx 1$ and for sparse systems $\nu<1$,
we expect that $\epsilon_h$ will take larger values than  $\epsilon_r$, especially for increasing  system size (increasing $M$, $N$).

We see that $D_{ij}^{(3)}\propto \epsilon_r \epsilon_h^2$ while $D_{ij}^{(4)}= C_1 \epsilon_h^4+C_2 \epsilon_h^2\epsilon_r^2$ with $C_1,C_2$  coefficients independent of $\epsilon_r, \epsilon_h$.
Note that the second term of $D_{ij}^{(4)}$ is of strictly higher order than $D_{ij}^{(3)}$.
In contrast, the first term of $D_{ij}^{(4)}$ is of order $\epsilon_h^4$ which can be comparable to $D_{ij}^{(3)}\propto \epsilon_r \epsilon_h^2$ if 
the imbalance $\epsilon_r \ll \epsilon_h \ll 1$ becomes so large that $\epsilon_r$ and $\epsilon_h^2$ become comparable.
Since this may indeed occur in our system, we treat $D_{ij}^{(3)}$ and $D_{ij}^{(4)}$ on an equal footing in the following, 
but discard the suppressed term $\propto \epsilon_h^2\epsilon_r^2$ in $D_{ij}^{(4)}$, i.e. we focus on
\begin{equation}
D_{ij}^{(4)}=-\frac{1}{12}\left(\frac{h}{2\pi R}\right)^4(u_{ij}^{(0)})^4. \label{d4eq2}
\end{equation}
Along the same lines, since the next order contribution to $D_{ij}$, $D_ {ij}^{(5)} \propto \epsilon_r \epsilon_h^4$ is of strictly higher order than both 
$D_ {ij}^{(3)}$ and $D_ {ij}^{(4)}$ it can be safely ignored in the current study.

The bending-induced corrections $D_ {ij}^{(3)},D_ {ij}^{(4)}$ to the interparticle distances immediately translate to corrections of the interaction energy 
of the bent helical segment $E_b=\frac{1}{2}\sum_{\substack{i,j=1 \\ i\neq j}}^N d_{ij}^{-1}$. Subtracting the energy $E_a$ of the corresponding straight helix as in Eq. (\ref{pothe1}) 
results in $\Delta E = E_b - E_a \approx \Delta E^{(3)}+\Delta E^{(4)}$ with
\begin{equation}
 \Delta E^{(n)}=-\frac{\lambda R^2}{4}\sum_{\substack{i,j=1 \\ i\neq j}}^N \frac{D_{ij}^{(n)}}{\left( R D_{ij}^{(2)}\right)^3}, \quad n=3, 4, \nonumber
\end{equation}
that is, explicitly,
\begin{equation}
\Delta E^{(3)}= -a_1\sum_{\substack{i,j=1 \\ i\neq j}}^N\frac{\left(\cos u_i^{(0)}+\cos u_j^{(0)}\right)(u_{ij}^{(0)})^2}{\left[2r^2\left[1-\cos(u_{ij}^{(0)})\right]+\left(\frac{h u_{ij}^{(0)}}{2\pi}\right)^2\right]^{3/2}} \label{de3}
\end{equation}
and 
\begin{equation}
\Delta E^{(4)}= a_2\sum_{\substack{i,j=1 \\ i\neq j}}^N\frac{(u_{ij}^{(0)})^4}{\left[2r^2\left[1-\cos(u_{ij}^{(0)})\right]+\left(\frac{h u_{ij}^{(0)} }{2\pi}\right)^2\right]^{3/2}}, \label{de4}
\end{equation}
where $a_1=\frac{\lambda r h^2}{16\pi^2 R}$ and $a_2=\frac{\lambda h^4}{768  \pi^4 R^2}$. 

A striking difference between these terms is that $\Delta E^{(3)}$ can obtain both positive and negative values depending on the exact GS configuration
while $\Delta E^{(4)}$ is strictly positive. Also, as shown in the Appendix, $\Delta E^{(4)}$ increases in general
faster than $\Delta E^{(3)}$ with the system size: $\Delta E^{(4)} =\mathcal{O}(N)$, $\Delta E^{(3)} =\mathcal{O} \left(\log N\right)$. Thus as the system size becomes larger  $\Delta E^{(4)}$ is expected to cross the term $\Delta E^{(3)}$
and dominate $\Delta E$ at least in cases where the GS configuration is close to equidistant. A further factor determining the contribution of $\Delta E^{(3)},~\Delta E^{(4)}$
in $\Delta E$ is the filling $\nu$. Since $\Delta E^{(3)} \propto \nu^2$ whereas  $\Delta E^{(4)} \propto \nu$ (see Appendix) denser systems will favor the dominance
of $\Delta E^{(3)}$. Regarding the dependence of $\Delta E^{(3)},~\Delta E^{(4)}$ on the helix radius
 $r$ (or equivalently the parameter $f$), we observe that apart from the denominator 
common in $\Delta E^{(3)},~\Delta E^{(4)}$ (eqs. (\ref{de3}),(\ref{de4})), $\Delta E^{(4)}$ does not depend in 
any other way on $r$, in contrast to $\Delta E^{(3)}$ whose numerator is proportional to $r$ (through its coefficient $a_1\propto r$), making $\Delta E^{(3)}$ 
much more sensitive to changes of $r$ (or $f$) than $\Delta E^{(4)}$.

\begin{figure}[htbp]
\begin{center}
\includegraphics[width=0.75\columnwidth]{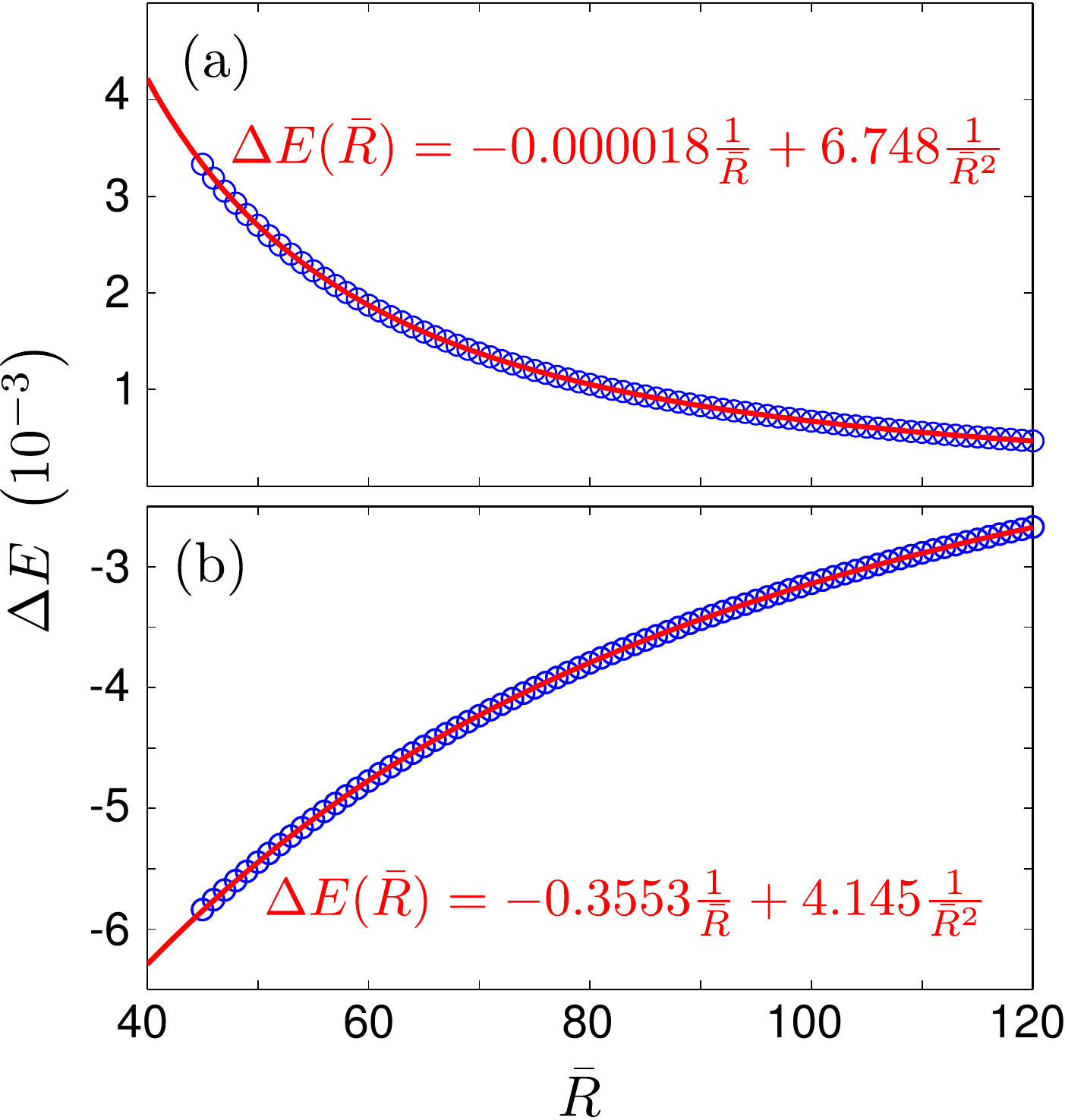}
\end{center}
\caption{\label{scalfig} (color online)  Scaling of the energy difference $\Delta E$ with the radius of curvature $R$
controlled by $\bar{R}$ (Eq. (\ref{Rcur})) for
$N=35$, $\nu=1$ and two different cases of the geometry parameter $f$: (a) $f=0$ (linear chain) and (b) $f=1$ (helical chain).
The blue circles are the result of numerical simulations and the red line is the fit
with Eq.~(\ref{scE1}). With this fit the coefficients $\bar{a}_1,~\bar{a}_2$ can be identified reading for (a) $\bar{a}_1=-0.000018 \approx 0$,
$\bar{a}_2=6.748$ and for (b) $\bar{a}_1=-0.3553$, $\bar{a}_2=4.145$.}
\end{figure}

The peculiar absence of a clear ordering of the Taylor terms $\Delta E^{(3)}$, $\Delta E^{(4)}$ contributing to the energy shift $\Delta E$ due to an infinitesimal
bending ($R \rightarrow \infty$) 
makes it difficult to define response parameters such as the bending rigidity $\kappa$ and the persistence length $l_p$ as typically used in the literature \cite{Balaeff2006,Nguyen1999, Han2003,Kiani2015,Trizac2016}. 
In particular, the definition of the bending rigidity requires a simple scaling of the energy difference
$\Delta E$ with the radius of curvature $R$ and the length of the system $L$. It turns out that in most systems such as e.g. in polyelectrolytes
modeled by uniformly charged cylinders \cite{Nguyen1999} or in magnetosome chains \cite{Kiani2015} $\Delta  E \propto LR^{-2}$, 
leading to a natural definition of the bending rigidity as
\begin{equation}
\kappa=\frac{2\Delta E R^2}{L}. \label{br1}
\end{equation}

In our case, however, the two relevant terms $\Delta E^{(3)}$, $\Delta E^{(4)}$ scale differently with $R$ and $L$, leading to a more involved scaling of $\Delta E$.
Particularly, we see from Eqs.~(\ref{de3}),(\ref{de4}) 
that $\Delta E^{(3)} \propto R^{-1}$ whereas $\Delta E^{(4)} \propto R^{-2}$ implying that in the response (large $R$) regime
\begin{equation}
\Delta E \approx a_1 \frac{1}{R}+a_2\frac{1}{R^2} ~~\textrm{or}~~ \Delta E \approx \bar{a}_1 \frac{1}{\bar{R}}+\bar{a}_2\frac{1}{\bar{R}^2}\label{scE1}
\end{equation}
with $a_1,a_2$ ($\bar{a}_1,~\bar{a}_2$) depending on all other system parameters except for $R$ ($\bar{R}$). 
The dependence on the length $L$ is even
more difficult to extract since it requires first a calculation of the sums in eqs. (\ref{de3}),(\ref{de4}). 
The only case for which this is possible is the limiting case $f=0$ corresponding to the linear chain. For this  
case $\Delta E^{(3)}=0$ (\ref{de3})) and it can also be shown that $\Delta E^{(4)}\propto L$. Thus only for the linear chain 
we recover the scaling $\Delta E \approx \Delta E^{(4)} \propto LR^{-2}$ allowing for a definition 
of the bending rigidity in terms of Eq. (\ref{br1}).

A representative example highlighting the different scalings of $\Delta E$ with $\bar R$
(or, equivalently, with $R$)
for two different values of the geometry parameter $f$ is shown in Fig.~\ref{scalfig}. For $f=0$ (Fig.~\ref{scalfig}(a))
we observe that $\Delta E \propto \bar{R}^{-2}$ since $\bar{a}_1\approx 0$, whereas for $f=1$ (Fig.~\ref{scalfig}(b)) we observe the mixed scaling law of Eq.~(\ref{scE1}),
which illustrates and confirms our above discussion.

To obtain the data underlying Fig.~\ref{scalfig}, $\Delta E$ has been calculated numerically by computing the energies $E_a$ and $E_b$
of the exact GS configurations of charges in the straight and the slightly bent helix, respectively. 
All relevant quantities have been calculated and presented in terms of dimensionless units provided  by scaling  position $x$ and energy $E$ with $\lambda$, $m$ and $2h/\pi$  as follows 
 \begin{equation}
  \tilde{x}=\frac{x \pi}{2h},~ \tilde{E}=\frac{2E h }{\lambda \pi},~\tilde{m}=1,~\tilde{\lambda}=1. \label{units1}
 \end{equation}
The same dimensionless units will be used in the following with the tilde being omitted for simplicity.

We proceed next to a case-by-case study of the  bending response of the helical chain of charges for different system parameters. 
For the sake of comparison we will use the same value $\bar{R}=50$ if not stated otherwise. 
Fig.~\ref{scalfig} confirms that for this value the system is well in the large-$R$ response regime characterized by the scaling of Eq.~(\ref{scE1}), at least for moderate numbers of particles.


 \begin{center}
 { \textbf{IV. BENDING RESPONSE WITH VARYING PARAMETERS}}
\end{center}

The aim of this section  is to analyze numerically the bending response of a number of charges confined on a helical segment
for different values of the number of particles $N$, the filling $\nu$ and the geometry parameter $f=\frac{2\pi r}{h}$.
As already mentioned, the most important step in the calculation of the energy difference $\Delta E$ characterizing the bending response 
is the identification of the GS configuration of the system. 

\begin{figure}[htbp]
\begin{center}
\includegraphics[width=0.88\columnwidth]{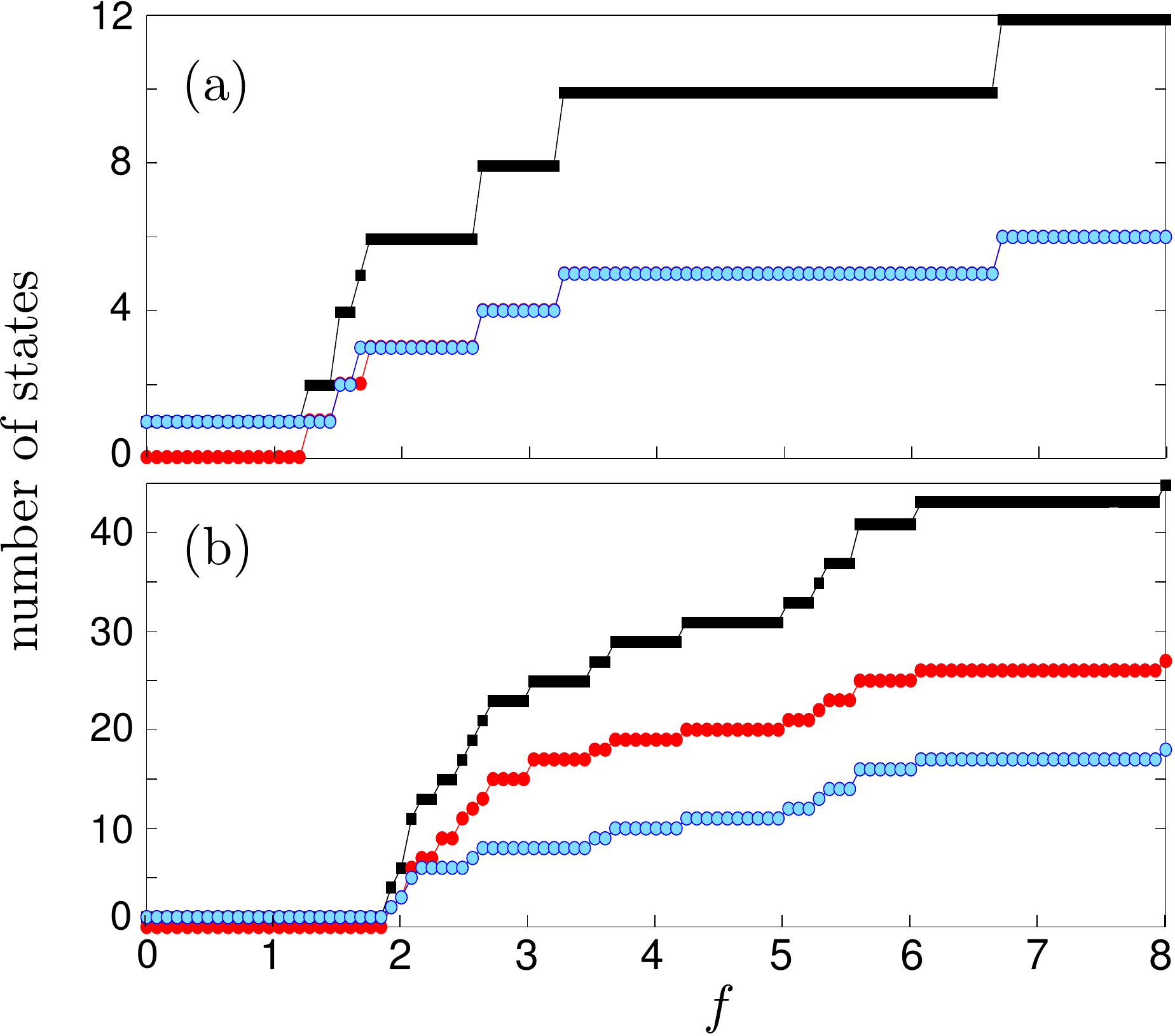}
\end{center}
\caption{\label{eqstates2} (color online)  Number of equilibrium states (black squares correspond to the total number of equilibrium states, red circles to the number of unstable states, and light blue (gray) circles
to the number of stable states)
as a function of the parameter $f$ for $N=4$ particles and  (a)  filling $\nu=1$ ($M=3$ windings), (b) filling $\nu=1/2$ ($M=6$ windings).}
\end{figure}

Identifying such a GS is a computationally hard task, becoming even intractable for large number of particles and large values of $f$ due to the complexity of the
potential energy landscape \cite{Schmelcher2011,Zampetaki2015a,Zampetaki2017} and the long-range character of the interactions. 
A very brief taste of this fact can be given already by inspecting the number of 
equilibrium states (both stable and unstable) as a function of $f$ for a very small system ($N=4$, Fig. \ref{eqstates2} (a),(b)). For small values of the parameter $f$ 
(starting from the straight line limit $f=0$) the system possesses only one state which is stable. Beyond some value $1<f<2$ more equilibrium states (both stable and unstable)
emerge, whose number quickly increases with $f$. This generic behaviour holds for different fillings (compare Fig. \ref{eqstates2} (a),(b)) but the partition of the states
into stable and unstable can be quite different due to the various possible bifurcations (e.g saddle-node or pitchfork)
through which new equilibria emerge \cite{Zampetaki2015a,Plettenberg2016,Zampetaki2017}.

Due to the complexity of even the smallest systems at larger values of $f$, we mainly focus here on the analysis  of the bending response of a helical chain of charges for small 
$f\lesssim 2$, for which the system possesses  a single equilibrium (its GS), leaving a brief discussion of the bending response of certain small systems for larger values of $f$ for the next section.

 \begin{center}
 { \textbf{ Bending response for the case $f \lesssim 2$}}
\end{center}

In the small-$f$ regime ($f\lesssim 2$) the GS of a helical chain of charges both for the straight and the bent helix
can be calculated relatively easily using as an initial guess the equidistant configuration and employing a 
standard constrained minimization method (e.g. conjugate gradient).  It is feasible therefore to study 
the energy difference $\Delta E$ between these GSs, characterizing the bending response, for different values of $f$ and different fillings $\nu$ 
for particle numbers ranging from $N=4$ to $N=100$. 

 \begin{center}
 { {A.} \textit{Commensurate fillings}}
\end{center}

\begin{figure}[htbp]
\begin{center}
\includegraphics[width=1\columnwidth]{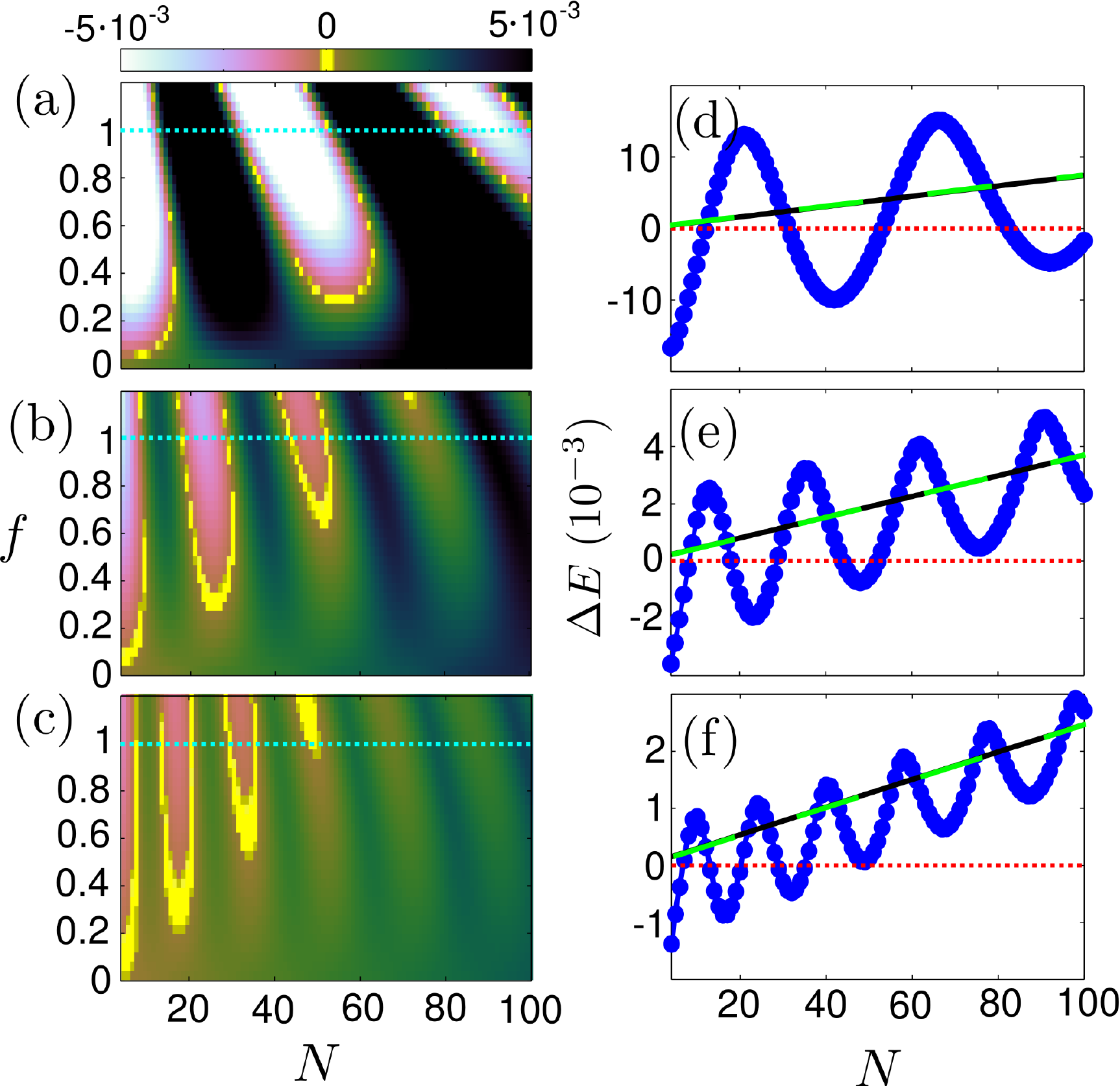}
\end{center}
\caption{\label{be_ri123} (color online) (a)-(c) Dependence of the bent-vs.-unbent GS energy difference $\Delta E$ of a helical chain of charges (depicted by color)
on the geometry parameter $f$ and the number of particles $N$ for decreasing commensurate fillings (a) $\nu=1$, (b) $\nu=1/2$, (c) $\nu=1/3$. The yellow color marks the values 
of $\Delta E \approx 0$. The horizontal cyan dotted line marks the case $f=1$ used for comparing to the linear case ($f=0$) in the panels of the right column. 
(d)-(f) $\Delta E$ as a function of $N$ for $f=1$ (blue line with dots, oscillatory) and $f=0$ (black line, linearly increasing) 
for $\nu=1,1/2$ and $1/3$ respectively. The green dashed line on top of the black depicts the values of $\Delta E^{(4)}$ for $f=1$
whereas the horizontal red dotted line marks the position of zero.}
\end{figure}

We first focus on the case of commensurate fillings, i.e. $\nu=(N-1)/M=1/n,~n=1,2,3, \ldots.$, found to exhibit particularly interesting properties. 
The dependence of the bent-vs.-unbent energy difference $\Delta E$ of a helical chain of charges on the particle number $N$ and the geometry parameter $f$ is shown in 
Fig.~\ref{be_ri123} (a)-(f) for the commensurate cases $\nu=1, 1/2$ and $1/3$.

For $f=0$ and every filling $\nu$, $\Delta E$ is strictly positive and increases linearly with the number of particles $N$ (Fig. \ref{be_ri123} (d)-(f)).
This is the expected result in the limiting case 
of a linear chain of charges since an arbitrarily small bending 
leads to smaller Euclidean interparticle distances and therefore larger potential energies, making the bending 
unfavorable for the system. More precisely, we have seen that to leading order
$\Delta E$ is comprised of two parts: $\Delta E^{(3)}$ (Eq. (\ref{de3})) and $\Delta E^{(4)}$ (Eq. (\ref{de4})). Since the former is zero for $f=0$, we have that 
$\Delta E \approx \Delta E^{(4)}$ which as shown in the Appendix increases linearly with $N$ ($\Delta E^{(4)} \propto N$, see Fig. \ref{be_ri123} (d)-(f))
 leading among others to the conclusion $\Delta E \rightarrow \infty$ for $N\rightarrow \infty$, well known for 1D Wigner crystals \cite{Nguyen1999}.

As $f$ gets larger, we observe an intriguing oscillatory behaviour of $\Delta E$ with the number of particles $N$ (Fig. \ref{be_ri123} (a)-(c)),
yielding also, especially for larger fillings (Fig. \ref{be_ri123} (a)), alternating signs of $\Delta E$. The oscillation period of $\Delta E$ 
decreases with the filling almost by a factor of $\nu$, leading to faster and
faster oscillations with respect to the parameter $N$ (Fig. \ref{be_ri123} (d)-(f)). Such an oscillating behaviour
of a quantity with the system size for fixed filling is rather unusual.
Essentially it means that the system does not behave universally when scaled, i.e. a larger version (both in terms of 
 the number of particles $N$ and the length $L$) of a particular system can have properties entirely different  from the initial 
system, favoring for example bending when the former does not.

The oscillations in $\Delta E$ are accompanied by a monotonous increment of their mean value (drift), so that for large enough systems 
$\Delta E$ is  overall positive. As shown in Fig.~\ref{be_ri123} (d)-(f) this drift seems to follow the behavior of the $\Delta E^{(4)}$
term (green dashed line), i.e. it is linearly increasing with $N$, pointing to the dominance in $\Delta E$ of the strictly positive $\Delta E^{(4)}$ for large $N$, a fact already predicted in Sec. III (see also  Appendix). Note that the $\Delta E^{(4)}$ term depends only weakly on $f$  
(through its denominator, see Eq. (\ref{de4})) 
for the low $f$-values  addressed here and thus the $\Delta E^{(4)}$ contribution for arbitrary $f\lesssim 2$ almost coincides with the $\Delta E \approx \Delta E^{(4)}$ for $f=0$ (compare green dashed line and black line in Fig. ~\ref{be_ri123} (d)-(f)).

\begin{figure}[htbp]
\begin{center}
\includegraphics[width=1\columnwidth]{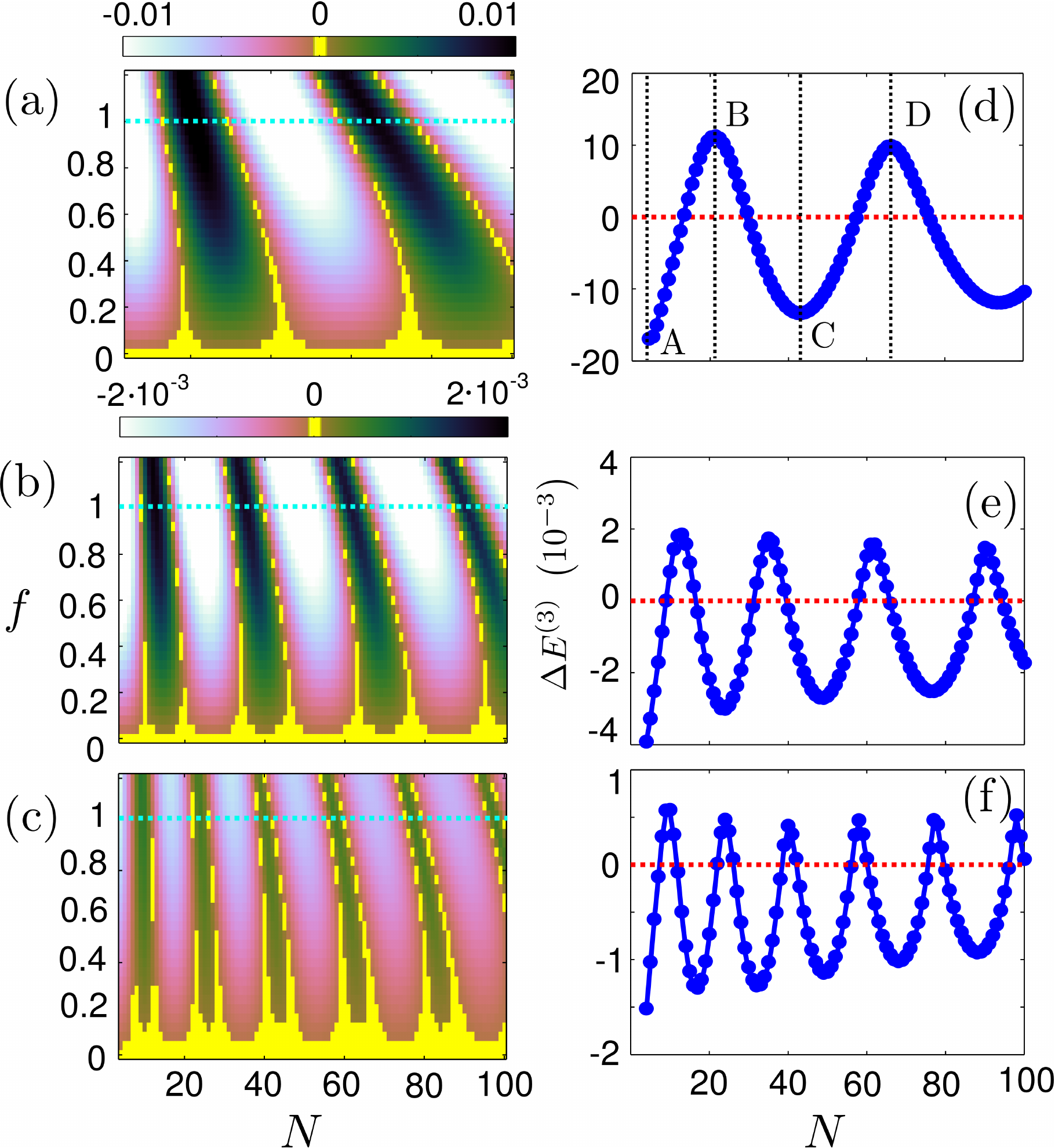}
\end{center}
\caption{\label{tay_123} (color online) (a)-(c) Dependence of  $\Delta E^{(3)}$ (depicted by color)
on the geometrical parameter $f$ and the number of particles $N$ for decreasing fillings (a) $\nu=1$,(b) $\nu=1/2$, (c) $\nu=1/3$. The yellow color marks the values 
of $\Delta E^{(3)} \approx 0$. The horizontal cyan dotted line marks the case $f=1$. (d)-(f) The term $\Delta E^{(3)}$ as a function of $N$ for $f=1$ (blue line with dots) for $\nu=1$, $1/2$ and $1/3$, respectively. 
The horizontal red dotted line marks the position of zero  and the vertical black dotted lines in (d) mark the positions of minima (A,C) and maxima (B,D) of $\Delta E^{(3)}$.}
\end{figure}

What remains to be investigated is the origin of the observed oscillations in $\Delta E$ and the dependence of their  period on the filling $\nu$.
Since these oscillations become more prominent for larger $f$ (or equivalently $r$), it is natural to assume that they have to do with the first contribution in $\Delta E$, i.e. with $\Delta E^{(3)}$.
Indeed, it turns out that $\Delta E^{(3)}$ (Eq.(\ref{de3})) shows an oscillating behaviour with the number of particles $N$ (Fig.~\ref{tay_123} (a)-(f)), very similar to that of Fig.~\ref{be_ri123} (a)-(f), save  the linear drift.
Again the frequency of the oscillations
increases with the filling almost by a factor $n=1/\nu$ and $\Delta E^{(3)}$ alternates between positive and negative values for $f\neq 0$. For $f=0$,  $\Delta E^{(3)}=0$ and the only contribution to $\Delta E$ comes from $\Delta E^{(4)}$ (Eq. (\ref{de4})) 
as discussed above.  

\begin{figure}[htbp]
\begin{center}
\includegraphics[width=0.85\columnwidth]{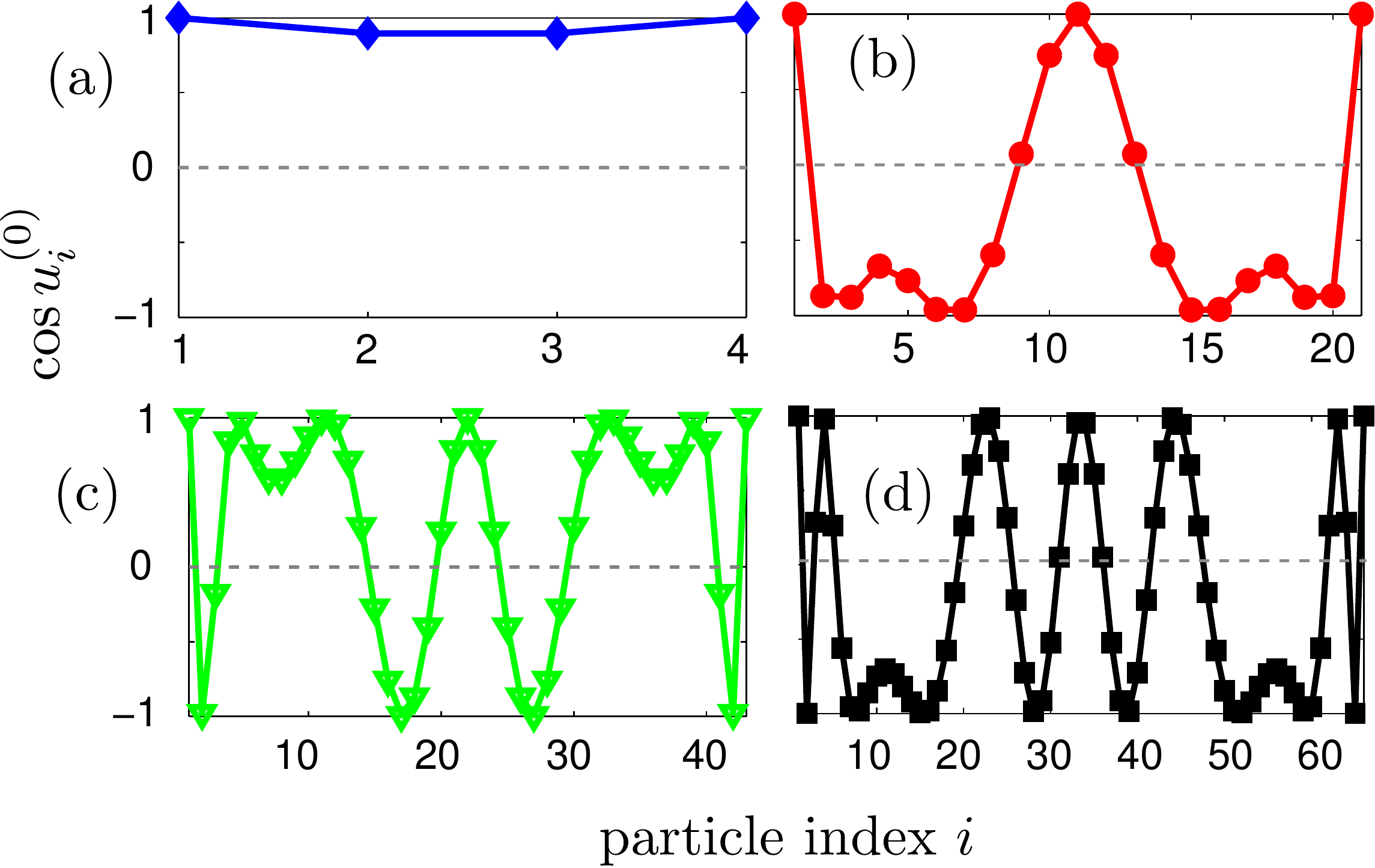}
\end{center}
\caption{\label{x_proj} (color online) (a)-(d) The $\cos u_i^{(0)}$ (determining the $x$-projection of the particles' position in the straight helix) as a function of
the particle index $i$ for (a) $N=4$, (b) $N=21$, (c) $N=43$ and (d) $N=66$ particles corresponding to the vertical lines (A,B,C,D) in Fig. \ref{tay_123} (d).
For all the cases $\nu=1$ and $f=1$ as in Fig.  \ref{tay_123} (d).}
\end{figure}

As can be seen from Eq.~(\ref{de3}), the sign of $\Delta E^{(3)}$ crucially depends on the cosines of the particles' angles $u_i$  which in view of  
Eq. (\ref{te1}) are proportional to  their $x$-projections. In Fig. \ref{x_proj} (a)-(d) we present
these projections for filling $\nu=1$, $f=1$ and four cases of particle numbers  $N$ corresponding to the minima and maxima of $\Delta E^{(3)}$ as shown in Fig. \ref{tay_123} (d) .

Obviously  $\cos u_i^{(0)}$ (Fig. \ref{x_proj} (a)-(d)) possesses an oscillatory profile (with respect to the particle index $i$)  with a number of maxima that increases for increasing $N$.
In particular, starting from $2$ maxima in
Fig. \ref{x_proj} (a) we go successively to $5$ in (b), $7$ in (c) and finally $9$ in (d). The outer particles reside as expected always in the edges 
$0$ and  $2\pi M$, a fact reflected by their cosine being always equal to $1$.  For the other particles the $x$-projection oscillates with the particle index $i$,
but the oscillations are not always full (i.e. do not always go from $-1$ to $+1$). Instead some maxima (minima) may be located at relatively low (high) values leading to extended regions of strictly negative (positive) $x$-projections.
These result in  the average of $\cos u_i^{(0)}$ with respect to  $i$ being positive for the cases of Fig. \ref{x_proj} (a),(c) corresponding to the minima of $\Delta E^{(3)}$ (note the global minus sign in Eq.~(\ref{de3}))
and  negative for the cases of Fig. \ref{x_proj} (b),(d) corresponding to the (positive) maxima of $\Delta E^{(3)}$. 

In other words, depending on the number of particles $N$ the helical chain 
can consist of more charges with positive $x$-projections or more particles with negative $x$-projections. Given that up to now we have assumed a bending of the helical chain
towards the positive $x$-direction (Fig. \ref{syst1}, Eq. (\ref{te2})), the former case corresponds to more particles occupying the outer side of the helix,
becoming stretched as a result of the bending,  
whereas the latter case corresponds
to more particles occupying the inner side of the helix which becomes compressed due to bending. It is this particle imbalance $\Delta N=N_O-N_I$ between 
the number of particles lying on the outer part of the helix $N_O$  and the
number of particles on the inner part $N_I$ which determines eventually the sign of $\Delta E^{(3)}$. If $\Delta N$ is positive,
 $\Delta E^{(3)}<0$ whereas  $\Delta N<0$ yields  $\Delta E^{(3)}>0$. But why does the particle imbalance $\Delta N$ between the outer 
 and the inner side oscillate with $N$? 

\begin{figure}[htbp]
\begin{center}
\includegraphics[width=0.85\columnwidth]{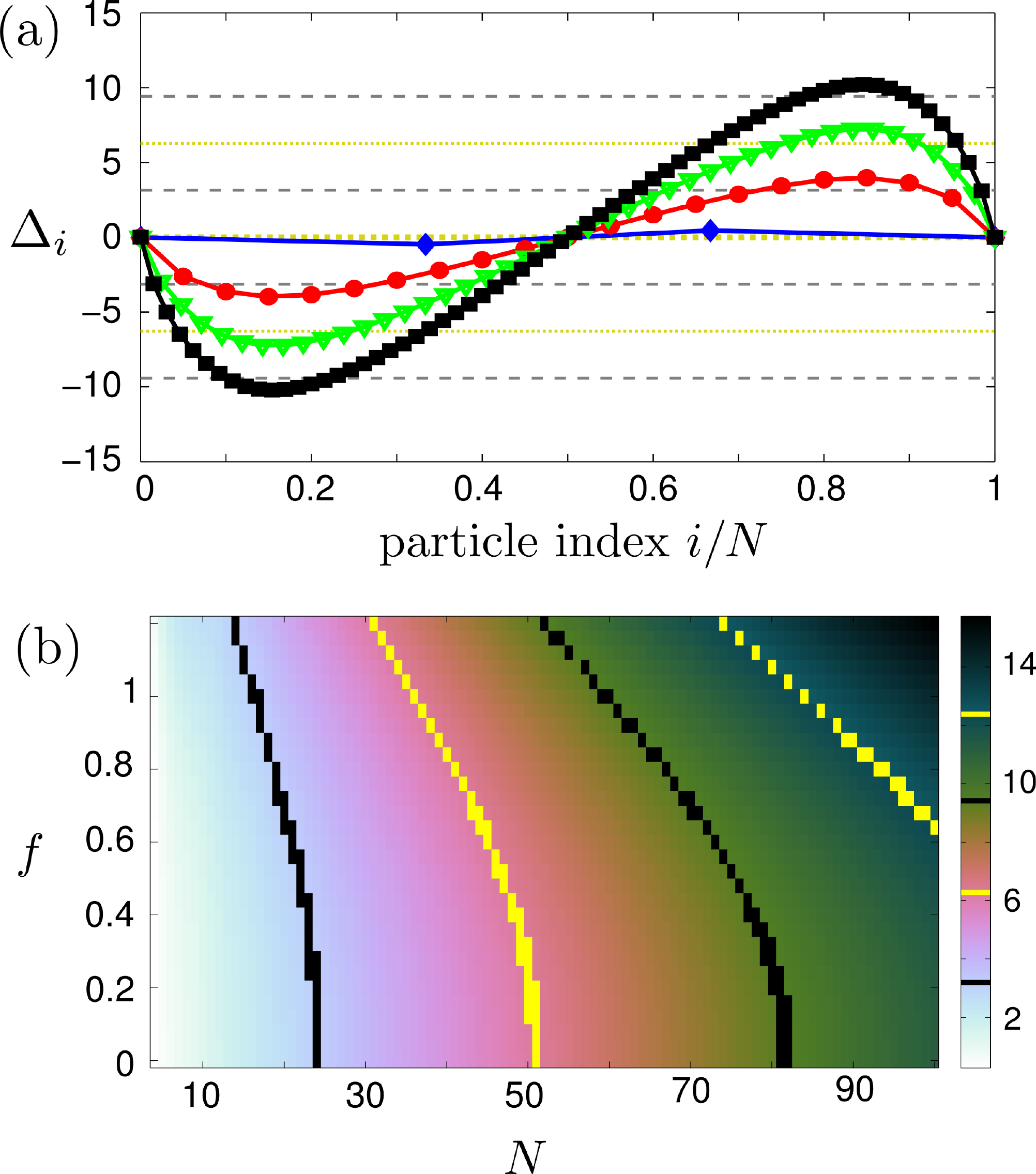}
\end{center}
\caption{\label{x_proj2} (color online)  (a) Dependence of  $\Delta_i=u_i^{0}-u_i^{eq}$ 
on the normalized particle index $i/N$ for the cases of Fig. \ref{x_proj} ($N=4$-blue line with diamonds, $N=21$-red line with circles, $N=43$-green line with triangles and 
$N=66$-black line with squares). The dotted dark yellow lines mark the even multiples of $\pi$ ($-2\pi,0,2\pi$) whereas the dashed gray lines the odd multiples ($-3\pi,-\pi,\pi,3\pi$). 
(b) The amplitude of $\Delta_i$,  $K=\max\left(\Delta_i\right)$, (depicted by color) as a function of the number of particles $N$ and the geometry parameter $f$. The yellow (bright)
lines in the spectrum mark the values of $K\approx 2 m\pi, m=1,2$, whereas for the black (dark) ones  $K\approx (2 m-1)\pi, m=1,2$.}
\end{figure}
 
In the studied case of $\nu=1$, for small $f\lesssim 2,
$ if the GS configuration of the helical chain consisted of the charges occupying equidistant positions ${u}_j^{eq}=2\pi (j-1)$
(including also the two edges $u_1^{eq}=0,~u_N^{eq}=N$ of the helical segment) \cite{Note2},
then all the charged particles would reside on the outer side of the helix with positive $x$-projections  (Fig. \ref{syst1}), since $\cos u_j^{eq}=1$ independently of $N$,
leading to strictly negative $\Delta E^{(3)}$ values
increasing in magnitude with $N$ (see Appendix). 
Due to the long-range character of the interactions and the fixed-ends boundary conditions imposed on the helical chain, the actual GS differs from the equidistant one.
This is already known \cite{Zampetaki2013a} for the case of a finite linear chain of ions (ions in the box) corresponding to our $f=0$ case. In particular, 
it has been shown \cite{Zampetaki2013a}
that the deviation from the equidistant configuration $\Delta_i=u_i^{(0)}-u_i^{eq}$ as a function of the particle index $i$
has a form similar to a skew sine with a minimum (maximum) around $N/5$ ($4N/5$). This is  the case for our system also for $f=1$
as depicted in  Fig. \ref{x_proj2}(a). The profile of $\Delta_i$, reminiscent of a skew sine, 
is overall the same for all particle numbers $N$, differing only in its amplitude $K=\max\left(\Delta_i\right)$ which increases with $N$.

It turns out that  this amplitude  $K$ increases with $N$ (Fig. \ref{x_proj2} (b)), passing successively through the values $0, \pi,2\pi, 3\pi,3\pi,4\pi \ldots$
Meanwhile due to the functional form of $\Delta_i$ with respect to $i$ (Fig. \ref{x_proj2} (a)) many of the $\Delta_i$ values accumulate close to the maximum/minimum $\pm K$,
whereas the remaining values are spread in the interval $(-K,K)$. Thus, if $K$ is close to an even multiple of $\pi$, the $\Delta_i$ values close to the maximum/minimum will be close to $\pm 2\pi m$
($m=1,2,3,\ldots$) whereas the rest will be spread in $(-2\pi m, 2\pi m)$.
Since $\cos u_i^{(0)}=\cos\left(u_i^{eq}+\Delta_i\right)=\cos \Delta_i$, the accumulation of $\Delta_i$ values close to $\pm K=\pm 2\pi m$ implies
that there will be an overall excess of particles with positive $x$-projections (positive $\cos u_i^{(0)}$), leading to $\Delta N>0$ (Fig. \ref{x_proj} (a),(c))
and consequently negative $\Delta E^{(3)}$ values.
In the opposite case $K\approx (2m-1)\pi$ the $x$-projections would be mostly negative (since the $\cos u_i^{(0)}$ accumulate around $\cos(\pm K)\approx-1$), i.e. $\Delta N<0$  (Fig. \ref{x_proj} (b),(d)), resulting in  positive $\Delta E^{(3)}$
values. It is this increasing deformation of the GS from the equidistant configuration with increasing particle number $N$
whose interplay with the periodicity of the helical segment leads to an oscillatory particle imbalance $\Delta N$ and in turn to oscillations of $\Delta E^{(3)}$ and 
$\Delta E$ with $N$.

In the low $f$ regime, the other cases of commensurate fillings $\nu=1/2,1/3,\ldots$  map approximately to the $\nu=1$ case by appropriately scaling  the deviations 
$\Delta_i \big|_{\nu=1}$ of the $\nu=1$ case. 
For the case of the linear chain  and a given number of particles $N$ the equilibrium particle positions are proportional to the system's length $L$ \cite{Zampetaki2013a}. 
In our case, for $f=0$ this can be interpreted as the equilibrium angles $u_i^{(0)}$ being proportional to the angular length $\Phi$. Obviously, 
also the equidistant angles $u_i^{eq}=\frac{\Phi(i-1)}{N-1}$ are proportional to $\Phi$ yielding
\begin{equation}
\Delta_i \propto \Phi=2\pi M=2\pi\frac{1}{\nu} (N-1) \label{del_i}
\end{equation}
 Therefore,  for $f=0$ and a particular value of $N$ we have $\Delta_i\big|_{\nu}=\frac{1}{\nu} \Delta_i\big|_{\nu=1}$.
For non-zero  values $f\lesssim 2$  the helical chain is close to the linear one and it possesses very similar GS structures as indicated by the form of the $\Delta_i$ profiles
(Fig. \ref{x_proj2} (a)). Thus, in this low $f$ regime the aforementioned scalings for the $f=0$ case, and particularly
$\Delta_i\big|_{\nu}\approx\frac{1}{\nu} \Delta_i\big|_{\nu=1}$, approximately hold.
As a result also the maximum value $K$ of $\Delta_i$  scales approximately with $\frac{1}{\nu}$, changing almost $1/\nu$ times more rapidly from an even to an odd multiple of $2\pi$ as $N$ 
increases, causing faster oscillations with respect to $N$ in $\Delta E^{(3)}$,  roughly by a factor of $\frac{1}{\nu}$ (Fig. \ref{tay_123} (d)-(f)).

Summing up the above results, we see that the energy difference $\Delta E$ characterizing the bending response of our system 
consists, at commensurate fillings and small $f$, of a linear increment with $N$,  stemming from the $\Delta E^{(4)}$ contribution to $\Delta E$,
accompanied by oscillations in $N$ featured in $\Delta E^{(3)}$.
The latter originate from the long-range character of interactions (see also Sec. V.)  which in the presence of the fixed-boundary conditions causes 
a global modulation of the GS angle coordinates of the charges (compared to the equidistant configuration). When this smooth modulation is 
superimposed with  the helical confinement having a characteristic angle scale $2\pi$, it produces the observed filling-dependent oscillations 
of $\Delta E$ with the particle number.

\begin{figure}[htbp]
\begin{center}
\includegraphics[width=1\columnwidth]{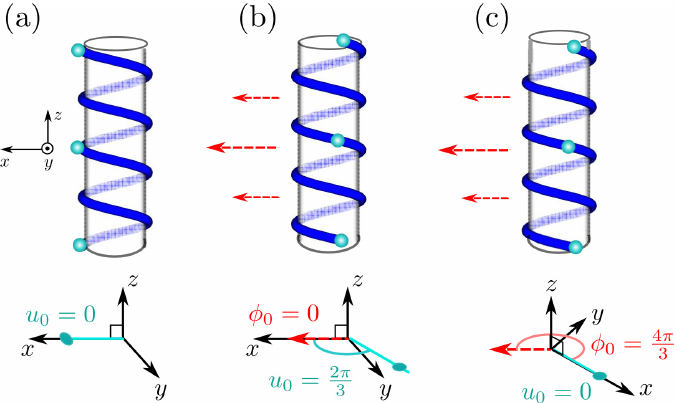}
\end{center}
\caption{\label{bend_explan1} (color online) (a) A charged helix with an offset parameter $u_0 = 0$ as in Fig. 1 (a).
(b) The same charged helix as in (a)  but with an offset parameter $u_0 =\frac{2\pi}{3}$,
which leads to a rotation (marked by the angle in the $x$-$y$
plane) around the helix axis. The bending direction for bending towards
$+x$ (corresponding to a bending angle $\phi_0=0$) is indicated by the red
dashed arrows. (c) The  charged helix of (a) with zero offset $u_0=0$
viewed from a different perspective. Bending it into the direction of
the bending angle $\phi_0 = 2\pi - \frac{2\pi}{3}=\frac{4\pi}{3}$ is equivalent to the bending shown in
(b).
}
\end{figure}

As discussed, the effect of the fixed boundaries on the bending response of our helical  chain of charges is substantial, since they determine the character of the GS configuration.
So far we have assumed 
that the edges of the helical segments are given in terms of angles by  $u_{in}=0$ and 
$u_{end}=2\pi M$ (both corresponding to $x=r,~y=0$, see Eq.~(\ref{te1}), Fig.~\ref{syst1} (a)). 
In general, we could obtain a helical segment with an integer number of windings $M$ 
if its endpoints satisfied $u_{in}=u_0$ and 
$u_{end}=2\pi M+u_0$, with an overall offset $u_0 \in[0, 2\pi)$.
The effect of $u_0$ is a combined
rotation around the $z$-axis by an angle $u_0$ and a translation along $z$ (compare Figs. \ref{bend_explan1}(a),(b)).
The translation is irrelevant to the bending response, but the rotation
is not since we always consider bending towards the positive $x$-axis
and the rotation angle $u_0$ controls the relative orientation between the
helix segment and that axis. As an example, for $u_0 = \pi$  we obtain a
helical chain mirror symmetric to the one  obtained for $u_0 = 0$ (as
presented in Fig. \ref{bend_explan1} (a)) with the  charges occupying the opposite side
of the helical segment compared to the $u_0 = 0$ case. This affects 
accordingly the values of $\Delta E$ with respect to bending along the
$+x$-direction. Note that there is no such directional dependence for a
simple linear chain of ions which is invariant under rotations around
its axis.

Instead of rotating the helix segment and then bending along $+x$ (see Fig. \ref{bend_explan1} (b)), one
can equivalently think of bending the original, unrotated helix segment
in a direction different from $+x$. Denoting the angle characterizing
the direction of bending, termed hereafter as the bending angle,  by $\phi_0$,
it turns out that the bending of a helical  chain with $u_0 = 0$ in the
$\phi_0$-direction (Fig. \ref{bend_explan1}(c)) is equivalent  to the bending in the $+x$-direction ($\phi_0 = 0$)
of a helical chain with a boundary offset $u_0 = 2\pi-\phi_0$  (Fig. \ref{bend_explan1} (b)). 
For example the bending towards the $\pm x$ direction corresponds to $u_0=2\pi-\phi_0=0,\pi$ whereas a bending towards the $\pm y$ to $u_0=2\pi-\phi_0=3\pi/2,\pi/2$.
Thus the different  values of $\phi_0$ affect the values of cosines similarly to $u_0$, that is 
$\cos (u_i^{(0)}) \rightarrow \cos (u_i^{(0)}-\phi_0)$, i.e. the different values of $\phi_0$ result in  a further modulation in $\Delta E^{(3)}$ (see Eq. (\ref{de3}))
and subsequently in the energy difference $\Delta E$.

\begin{figure}[htbp]
\begin{center}
\includegraphics[width=0.75\columnwidth]{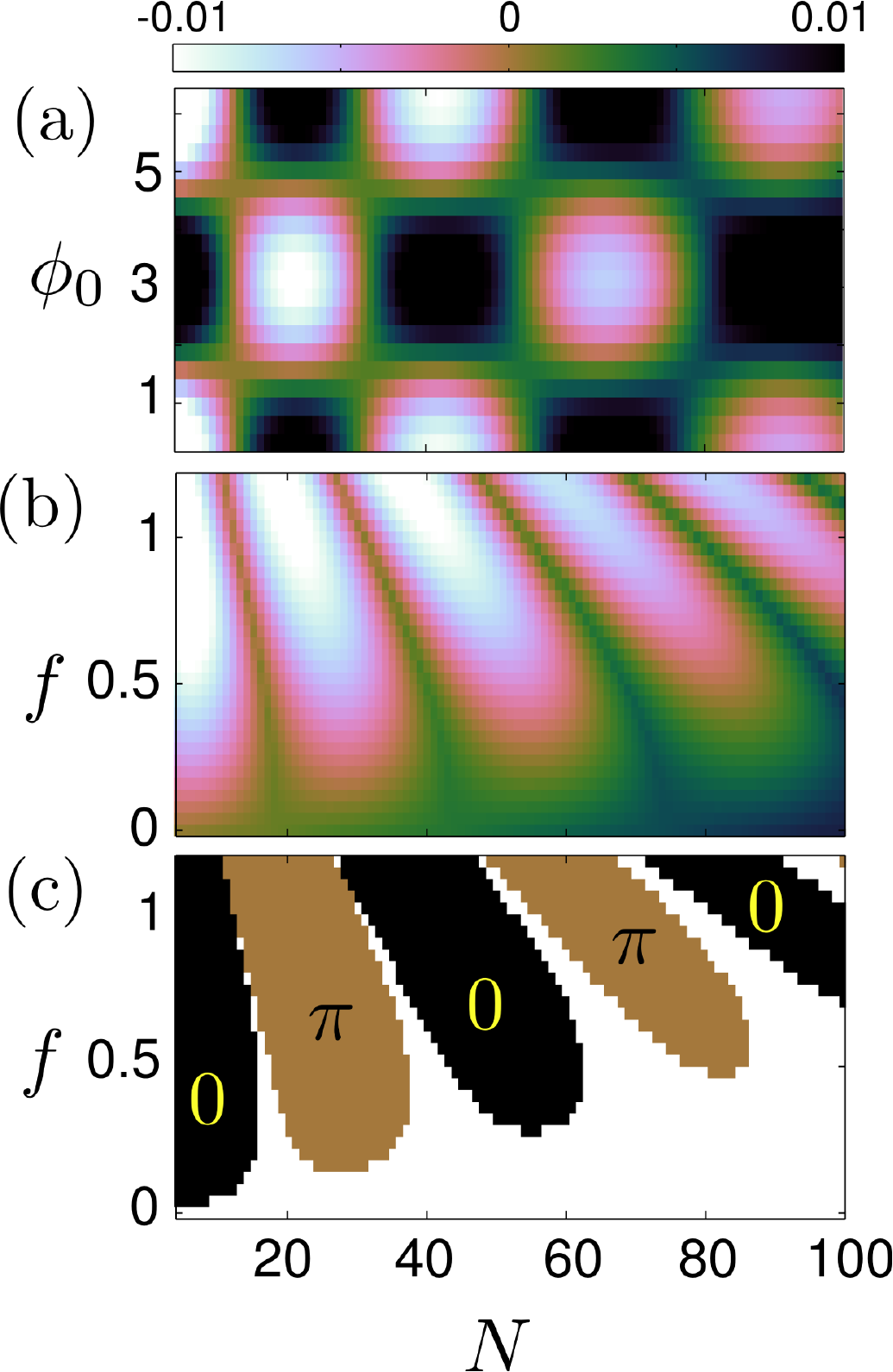}
\end{center}
\caption{\label{difends} (color online) The subfigures (a)-(c) present the effect of the bending angle $\phi_0$ on the bending response of
a commensurate helical chain of filling $\nu=1$ in the low-$f$ regime.
In particular we show
(a) the dependence of  $\Delta E$ (depicted by color) on the number of particles $N$ and the bending angle $\phi_0$ for $f=1$,
(b) the minimum value $\Delta E^*$ of $\Delta E$ (depicted by color) obtained by scanning $\phi_0 \in [0, 2\pi)$ as a function of the number of particles $N$ and
the geometrical parameter $f$ and (c) the preferable bending angles $\phi_0^*$  (depicted by color) corresponding to the most negative value $\Delta E^*$ (of subfigure (b)) as a function of 
the number of particles $N$ and the geometrical parameter $f$. The white region corresponds to regions of positive $\Delta E^*$ where bending in any direction is unfavorable. 
}
\end{figure}

Focusing on the behaviour of $\Delta E$ for the case of filling $\nu=1$ and $f=1$ (Fig. \ref{difends} (a)) we observe that apart from the oscillations with the particle number $N$  there are also oscillations with the bending angle 
$\phi_0$ providing a checkerboard-like pattern. This causes some regions with $\Delta E>0$ for $\phi_0=0$ to possess  $\Delta E<0$ for $\phi_0\approx \pi$, pointing to
the existence of an energetically preferable bending angle $\phi_0^*$ (in general non-zero) for each value of the particle number $N$.  Since we are interested in the general
bending response of our system it is necessary to take into account all the possibilities for bending. Along these lines, after scanning all the possible bending angles $\phi_0$
we have obtained the minimum energy difference $\Delta E ^{*}$ (corresponding to the optimal bending angle $\phi_0^*$) as a function of $N,f$. 

As shown in Fig. \ref{difends} (b) 
$\Delta E^*$ is mostly negative especially for larger values of $f$ and smaller particle numbers  $N$, indicating that the helical chain in this parameter regime in general favors bending into some direction.
Note that this also holds for certain regimes for which $\Delta E>0$ for $\phi_0=0$ (compare with Fig. \ref{be_ri123} (a)). It turns out (Fig. \ref{difends} (c)) 
that these regimes correspond to  parameter regions for which the preferable bending angle is $\phi_0^*=\pi$ instead of $\phi_0^*=0$, i.e. to cases where
the helical chain favors bending towards the $-x$ direction due to the charges occupying mostly the right part of the helix. Therefore, the oscillations in the values of 
$\Delta E$ for $\phi_0=0$ (Fig. \ref{be_ri123} (a)) are replaced with oscillations in the preferable bending angle $\phi_0^*$ from $0$ to $\pi$ with $N$ 
(Fig. \ref{difends} (c)). These oscillations stop for large enough values of $N$ since $\Delta E^*$ becomes eventually positive due to its domination by the positive monotonously
increasing $\Delta E^{(4)}$ term, as we have already discussed for the case of $\phi_0=0$.

Having highlighted the bending response of a helical chain of charges for the interesting case of commensurate fillings $\nu=1/n$ let us now
complete our discussion for the low $f$ regime  by a brief reference to the case of other fillings.

 \begin{center}
 { {B.} \textit{Other fillings $\nu$}}
\end{center}

For fillings other than $\nu=1/n$ the energy difference $\Delta E$, quantifying the bending response  of a helical chain of charges towards the $+x$ direction
($\phi_0=0$, Fig. \ref{syst1} (b)), mostly appears to have a smooth monotonous behaviour increasing with the number of particles $N$
and decreasing slightly with the geometry parameter $f$.

\begin{figure}[htbp]
\begin{center}
\includegraphics[width=1\columnwidth]{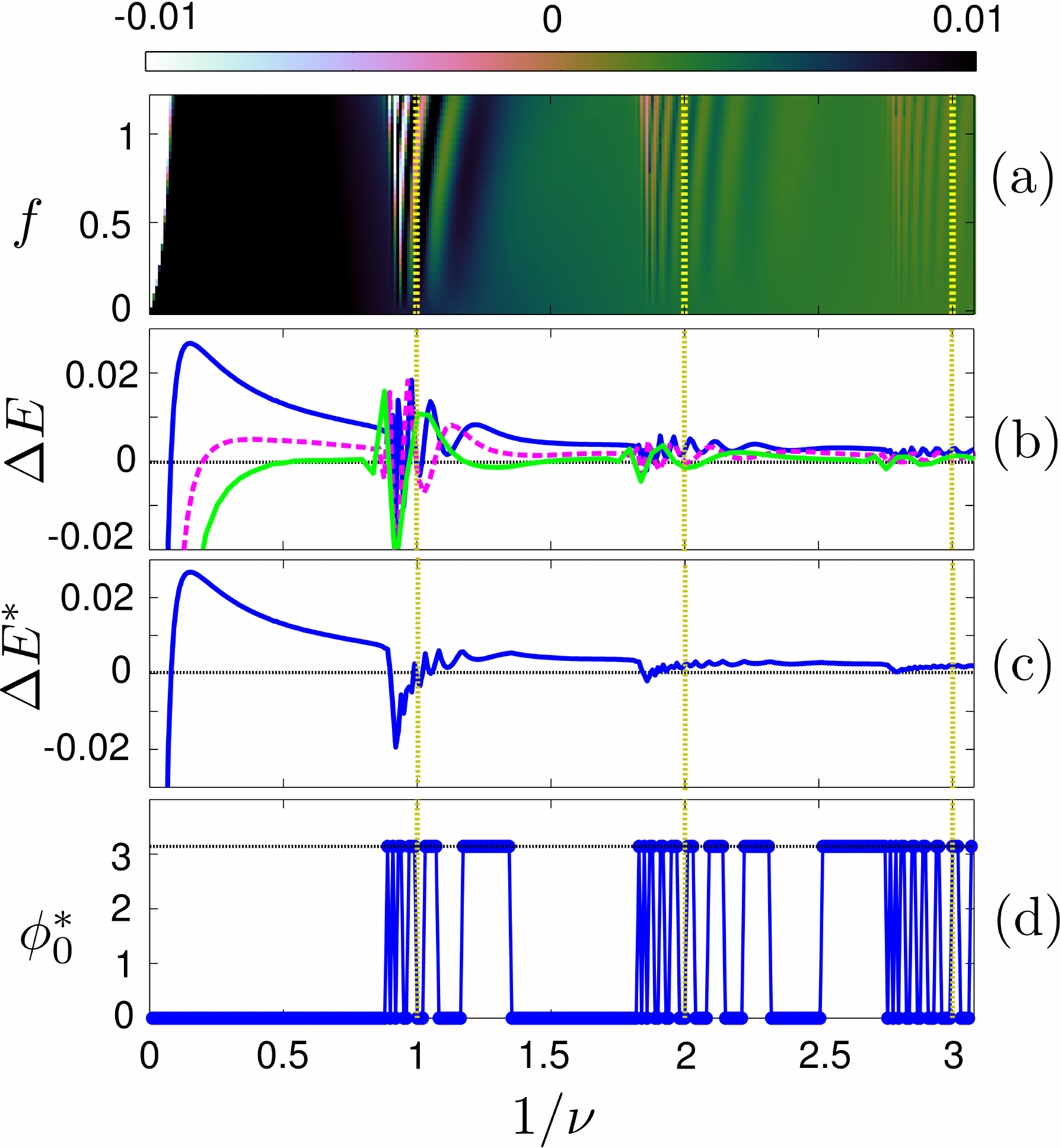}
\end{center}
\caption{\label{ofil2} (color online) 
(a) The energy difference $\Delta E$ (depicted by color) as a function of the inverse filling $1/{\nu}$ and the geometrical parameter $f$ for $N=100$ and $\phi_0=0$.
(b) The energy difference $\Delta E$ as a function of the inverse filling $1/{\nu}$ for  $\phi_0=0$, $f=1$ and for different number  of particles $N=100$ (blue (dark) solid line),
$N=50$ (magenta dashed line) and $N=25$ (light green (gray) solid line). Note that in the figure these appear successively  from top to bottom.
(c) $\Delta E^*$, the minimum value of $\Delta E$ resulting from a scan in $\phi_0 \in [0, 2\pi)$ as a function of the  inverse filling 
$1/{\nu}$ for $N=100$.
(d) The preferable bending angle $\phi_0^*$ corresponding to $\Delta E^*$ of subfigure (c)
as a function  of the inverse filling 
$1/{\nu}$ for $N=100$. The vertical dotted lines mark the first three positions of commensurate fillings $1/ \nu=1,2,3$  and the horizontal black dotted line marks the position of $\Delta E=0$
or $\phi_0^*=\pi$.}
\end{figure}

Such a smooth monotonous behaviour can be observed in  Fig. \ref{ofil2} (a) depicting  $\Delta E$  as a function of the parameter $f \lesssim 2$
and the inverse filling $1/\nu$ for  a fixed $N=100$. As shown, for most fillings $\Delta E$ possesses a high positive value decreasing with $1/\nu$ (i.e. increasing with $\nu$) (green background),
which is almost unaffected by the increment of the geometrical parameter $f$ (or equivalently the increase of the radius $r$). 
As discussed in more detail in the Appendix, this behaviour of $\Delta E$ points to it being  dominated, for most fillings, by the   positive $\Delta E^{(4)} \propto \nu$  which depends only very weakly on $r,f$ 
(it is independent of $r$ within the approximations of the Appendix). Indeed, even in the scenario of a highly ordered GS configuration close to equidistant $u_i^{(0)}\approx\frac{2\pi (i-1)}{\nu}$,
the contribution of
$\Delta E^{(3)}$ is expected to be suppressed for incommensurate fillings due to the highly scattered values of $\cos u_i^{(0)}$ for these cases,
leading to terms that cancel out after summation 
in Eq. (\ref{de3}). 
The $\Delta E^{(3)}$ contribution is expected to become
substantial only in cases for which the values  $\cos u_i^{(0)}$ cluster around a strictly positive (negative) value e.g. around $1$ (-$1$), i.e. only in cases of
very large ($\nu \gg 1$) or commensurate ($\nu=1/n$) fillings (see also the discussion in the Appendix).

Figure \ref{ofil2} (a) confirms exactly this behaviour.
As $f$ increases and for very large fillings ($1/\nu \rightarrow 0$) a small region of highly negative $\Delta E$ emerges  as a result of the dominance of the $\Delta E^{(3)}$ term. The most intriguing behaviour is
encountered for large enough values of $f$ and fillings close to the commensurate ($1/\nu=1,2,3,\ldots$). We observe that $\Delta E$ performs oscillations 
with the inverse filling
which are localized around the commensurate values and whose amplitude decreases as $1/\nu$ increases from one commensurate value to the next.
This behaviour is generic for different numbers of particles (Fig. \ref{ofil2} (b)) with 
a slight shift of the oscillations to lower values of $1/\nu$ and an overall
decrease of $\Delta E$ for decreasing $N$ (the latter resulting  from the fact that for incommensurate fillings $\Delta E \approx \Delta E^{(4)}$ which is approximately proportional to
$N$ ). 

The oscillations of $\Delta E$ with the inverse filling $1/\nu$ around $\nu=1,1/2,1/3$ are of the same origin as
its oscillations with $N$ we have encountered above (Fig. \ref{be_ri123}). 
If the charges in the GS configurations were equidistant, these oscillations would collapse to distinct dips positioned exactly at $1/\nu=1,2,3$, since at these fillings
each particle would attain the maximum $x$-projection and the contribution from $\Delta E^{(3)}$ would be maximal. Due to the fixed boundaries there is a substantial
deviation $\{\Delta_i\}$ of the GS configuration $\{u_i^{(0)}\}$ from the equidistant one $\{u_i^{eq}\}$ with a form of a skew sine (cf. Fig. \ref{x_proj2}(a)) for any filling.
For fillings close to commensurate ($1/\nu=1,2,3,\ldots$) we have  that $u_i^{eq}$ is approximately a multiple of $2 \pi$
leading to $\cos u_i^{(0)}\approx \cos \Delta_i$. As discussed in the previous subsection this leads to the values $\cos u_i^{(0)}$ accumulating around
$\cos (\pm K)$, with $K$ denoting the maximum of $\Delta_i$. It has been also argued (Eq. (\ref{del_i})) that $K$ is roughly proportional both to $1/\nu$ and the number of particles 
$N$. Thus for a large enough value of $N$ a slight  change of $1/\nu$ can lead to a change of $K$ from an even to an odd multiple of $\pi$, resulting in an
excess of positive or negative values of $\cos u_i^{(0)}$, respectively. This in turn implies that for close-to-commensurate values as   $1/\nu$ increases, $\Delta E^{(3)}$ will alternate between
negative and positive values, inducing the observed oscillations in $\Delta E$. Similarly to the context of the $N$-dependent oscillations for commensurate fillings, analyzed in the above subsection, 
 these regions of oscillations can be interpreted as parameter regimes with an overall small value (negative) of the minimum bending energy difference $\Delta E^*$
(Fig. \ref{ofil2} (c)) and the optimal bending angle $\phi_0^*$ switching from $0$ to $\pi$ and vice versa (Fig. \ref{ofil2} (d)).

These results complete the main part of our paper regarding the bending response of the helical chain of charges in the small-$f$ regime where it is relatively close to a linear chain.
In the next sections we present a number of results on extensions of the problem that we have not addressed so far, such as the effect of possible interaction screening, the bending response for 
larger values of $f \gtrsim 2$ and the beyond-response regime.


\begin{center}
 {V.} \textbf{EFFECTS OF SCREENED INTERACTIONS}
\end{center}
So far we have assumed that the charges interact through bare Coulomb interactions, see Eq. (\ref{pothe1}). In order to study the effect of 
the range of interactions on the bending response as shown above in Fig. \ref{be_ri123}, we have examined the quantity $\Delta E(f,~N)$ in the case of screened Coulomb interactions. The generalized interaction potential reads
\begin{equation}
 V(\{u_i\})=\frac{1}{2}\sum_{\substack{i,j=1 \\ i\neq j}}^N  \frac{\exp\left[-\alpha \abs{\vec{r}(u_i)-\vec{r}(u_j)}\right]}{\abs{\vec{r}(u_i)-\vec{r}(u_j)}},\label{scr1}
\end{equation}
with $\alpha^{-1}$  the screening length and the helical curve $\vec{r}(u)$ given by Eq. (\ref{te1}) for the unbent or  Eq. (\ref{te2}) for the bent conformation. 

\begin{figure}[htbp]
\begin{center}
\includegraphics[width=1\columnwidth]{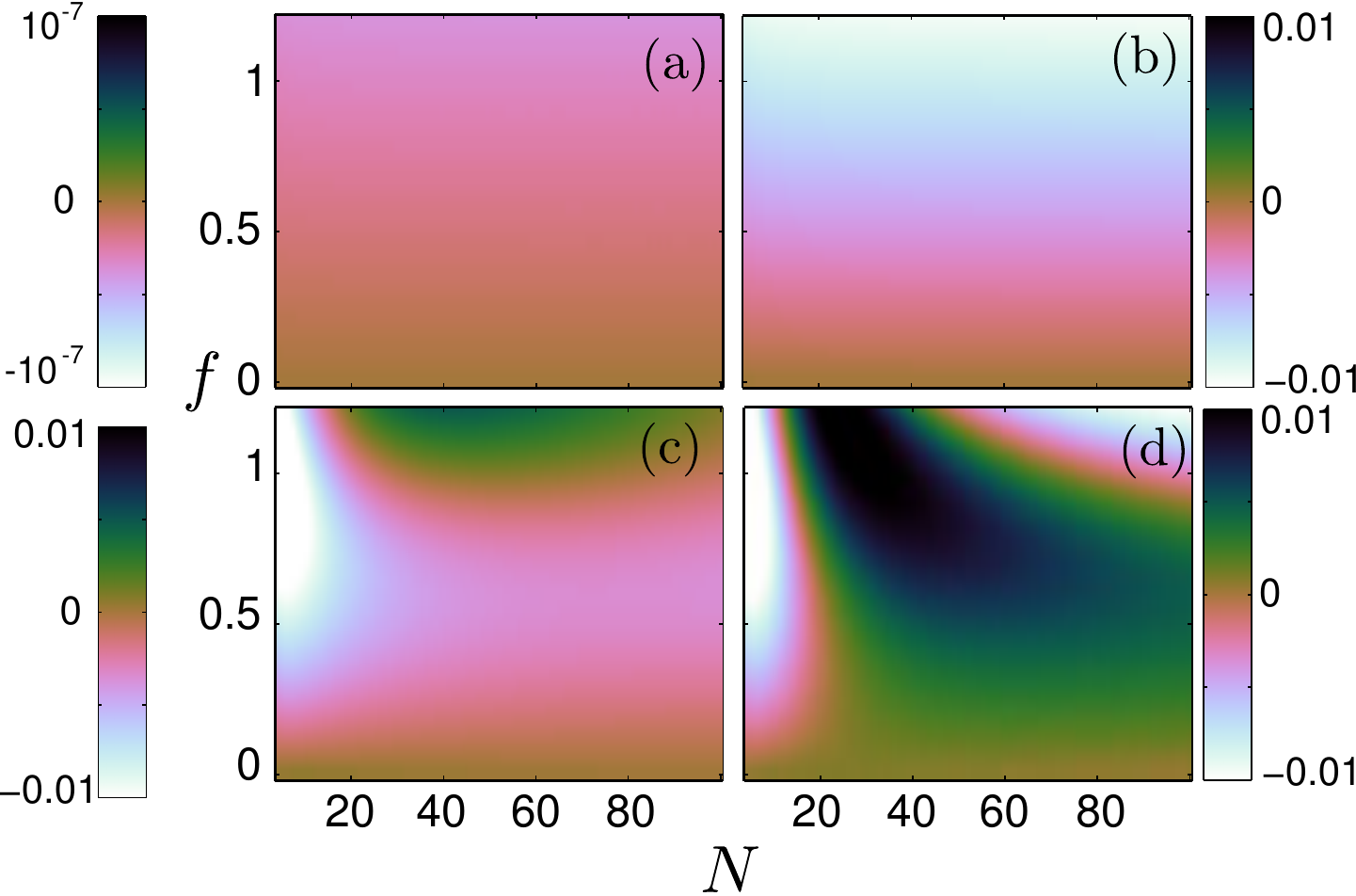}
\end{center}
\caption{\label{rsfig1} (color online) Dependence of the energy difference $\Delta E$ (depicted by color) of a helical chain of charges interacting through 
screened Coulomb interactions (Eq. (\ref{scr1})) on the geometrical parameter $f$ and the number of particles $N$ for filling $\nu=1$, bending angle $\phi_0=0$ and for different values of the screening length $a^{-1}$:
(a) $\alpha^{-1}=0.1$, (b) $\alpha^{-1}=1$, (c) $\alpha^{-1}=5$, (d) $\alpha^{-1}=10$.}
\end{figure}

Our results for the case of a commensurately filled ($\nu=1$) helical filament and for different values of the screening length $\alpha^{-1}$ 
are presented in Fig. \ref{rsfig1}. We observe
that  for interactions of a very short range (Fig. \ref{rsfig1} (a),(b)) 
$\Delta E$ is strictly negative, having overall small values due to the large screening. The values of $\Delta E$ decrease (becoming more negative) with the geometrical parameter $f$  
and are rather independent of the number of particles $N$.
These observations can be linked to the fact that for small screening lengths 
the particles reside approximately in the equidistant configuration ${u}_j^{eq}$, i.e. independently of $N$ they occupy the outer side of the helix and therefore favor bending in the direction of $\phi_0 = 0$
($+x$-direction, see Fig. \ref{syst1}).

\begin{figure*}[htbp]
\centering
\includegraphics[width=1.8\columnwidth]{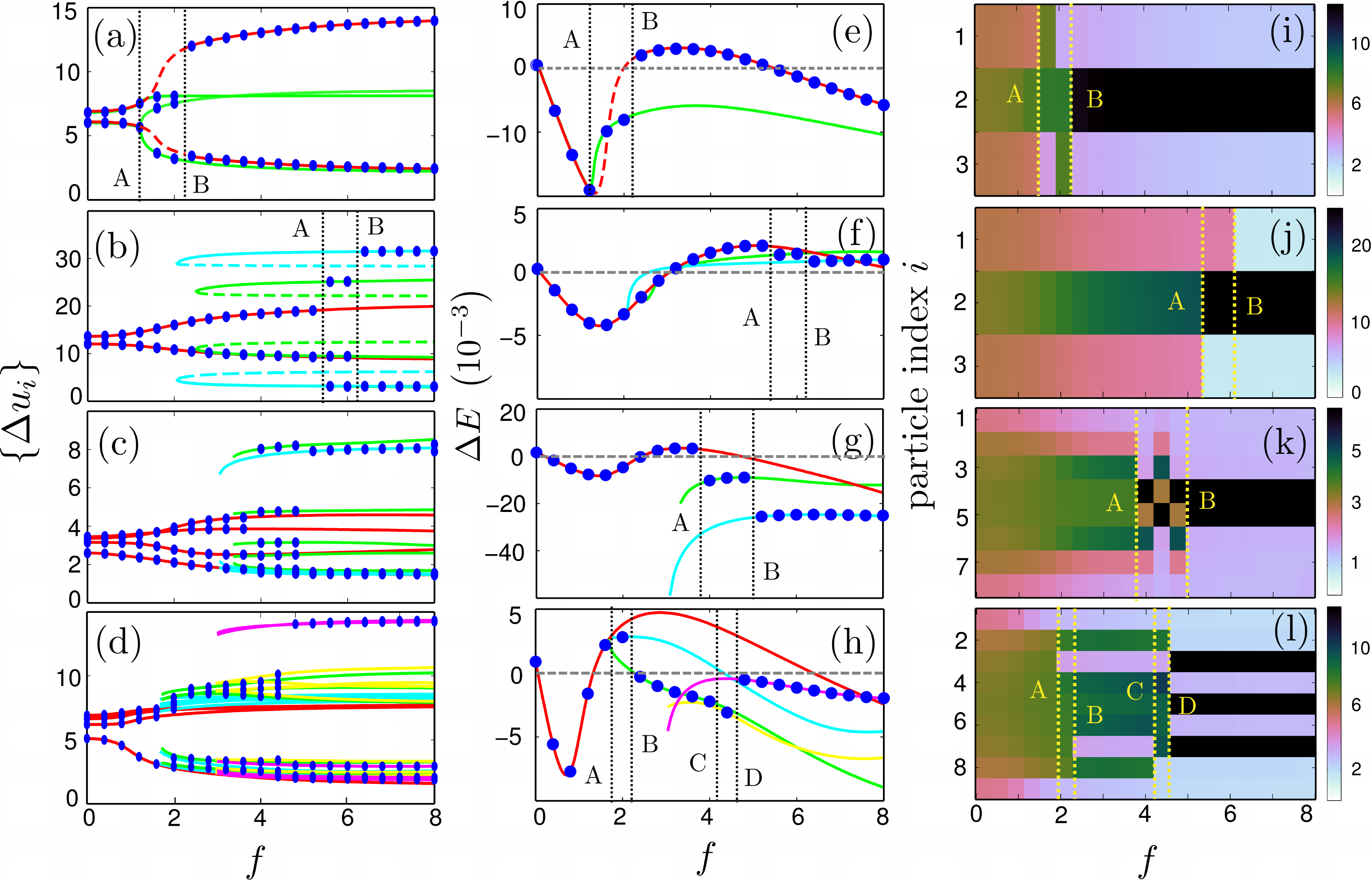}
\caption{\label{larger1} (color online) (a)-(d) Dependence of the interparticle distances  $\{\Delta u_i=u_{i+1}-u_i\}$ on the geometrical 
parameter $f$ for (a) $N=4,~\nu=1$, (b) $N=4,~\nu=1/2$, (c) $N=9,~\nu=2$, (d) $N=10,~\nu=1$. The blue circles mark the numerically obtained values for the GS,
whereas the colored solid lines depict the various solution branches (equilibrium configurations) relevant for the GS
configuration for different values of $f$. The dashed lines correspond to unstable solution branches.
All the branches that  appear without precursor correspond to the stable branches of saddle node bifurcations as in the case (b),
but their unstable pairs are not shown in Figs.~(c),(d) for reasons of clarity. Note that each branch (equilibrium state)
consists of $N-1$ interparticle distances $\Delta u_i$ some of which, due to symmetry, lie on top of each other.
(e)-(h) The energy difference  $\Delta E$ for $\phi_0=0$ as a function of $f$ for (e) $N=4,~\nu=1$, (f) $N=4,~\nu=1/2$, (g) $N=9,~\nu=2$, (h) $N=10,~\nu=1$
corresponding to the subfigures (a)-(d). Here also the blue circles stand for the numerically extracted GS $\Delta E$ values whereas the colored solid lines stand 
for the $\Delta E$ values corresponding to other equilibrium states whose interparticle distances are depicted in the subfigures (a)-(d) with the respective color.
(i)-(l) The  dependence of the GS $\Delta u_i^{(0)}$ values (depicted by color) on the particle index $i$ and the parameter $f$ for (i) $N=4,~\nu=1$, (j) $N=4,~\nu=1/2$, (k) $N=9,~\nu=2$, 
(l) $N=10,~\nu=1$. In all the cases the vertical dotted lines A,B,C or D correspond to the values of $f$ for which the energy of another branch crosses the previous GS energy.}
\end{figure*}

Only when the screening length is sufficiently increased, 
a region of positive $\Delta E$ emerges for large values of $f,N$ (Fig. \ref{rsfig1} (c)). This region moves to smaller $f,N$ values for larger interaction ranges
and another negative region of $\Delta E$ emerges (Fig. \ref{rsfig1} (d)) as a precursor of the oscillating pattern in the case of the infinite range of Coulomb interactions 
(Fig. \ref{be_ri123} (a)). We see therefore that the long-range character of the interactions is an essential feature
for the emergence of the oscillations of $\Delta E$ with respect to $N$ as presented in the previous section (Fig. \ref{be_ri123}).

Having discussed the effect of screening let us now have a look at the bending response of the helical chain of charges for larger values of the geometry parameter
$f$ than those examined in Sec. IV, allowing for much larger deviations of the helical chain from the linear one.

\begin{center}
{VI.} \textbf{BENDING RESPONSE FOR LARGER $f\gtrsim 2$}
\end{center}
We have seen that the system of helically confined charges in the low-$f$ regime ($f \lesssim 2$) possesses a single equilibrium
state, its GS  $\{u_i^{(0)}\}$, which changes smoothly with the geometrical parameter $f$. As already mentioned for larger values of $f$ the potential landscape of this 
system becomes more and more complex featuring more and more stable and unstable equilibria (Fig. \ref{eqstates2}). With varying $f$ these
equilibria form different fixed-point branches  (each branch consisting of the $N$ particles' angle coordinates $u_i$). 
Each one of the stable solutions for each $f$  constitutes a possible configuration $\{u_i\}$ in which the system can equilibrate and has its own bending tendency, characterized
by the energy difference $\Delta E$ between its bent and unbent conformation.

The stable equilibrium with the lowest potential energy for each $f$ corresponds to the respective GS of the system. Since the energies
of the different equilibrium branches might cross with increasing $f$, the GS in general does not follow a single branch but contains parts of different branches for different values 
of $f$. This results in the GS exhibiting dramatic changes in its structure as a function $f$ and subsequently  in its bending response, characterized by the GS energy difference $\Delta E$.
Some examples of this behaviour are shown in Fig. \ref{larger1} for a few cases with a small number of particles $N$.

In particular, we present the $f$-dependence of the $N-1$ interparticle distances $\{\Delta u_i=u_{i+1}-u_i\}$ (Fig.~\ref{larger1} (a)-(d)) 
and the corresponding bent-vs.-unbent energy difference $\Delta E$ (Fig.~\ref{larger1} (e)-(h)) for a number of different equilibria, each of which becomes the GS for certain values
of $f$. On top of them we also provide the corresponding GS values of $\{\Delta u_i^{(0)}\}$ and the respective GS $\Delta E$, complementing them with results about 
the $\Delta u_i^{(0)}$ profile as a function of $f$ (Fig. ~\ref{larger1} (i)-(l)).

As shown (Fig.~\ref{larger1}), these results vary vastly with $f$, $\nu$ and $N$, reflecting  the variety in the GS configurations. 
For small $f$, a single stable branch exists corresponding to the  GS which 
changes smoothly with $f$, as indicated by the smooth behaviour of the $N-1$ interparticle distances $\Delta u_i$ (Fig. \ref{larger1} (a)-(d)).
A common feature of the examined system in this regime is that the profile of  the GS $\Delta u_i^{(0)}$ is close to uniform, i.e. the interparticle distances are 
almost equal (Fig. \ref{larger1} (i)-(l)) except for  a small deviation caused by the fixed boundary conditions. As a consequence, the behaviour of the GS 
energy difference $\Delta E$ is smooth 
for increasing $f\lesssim 2$ (Fig. \ref{larger1}(e)-(h)), following the only existing stable branch, and is covered by our discussion in the previous section.

For larger values of $f$ the situation changes. More and more stable fixed-point branches corresponding to
equilibrium states with different energies emerge in this regime and for particular values of
$f$ one of these can become the new GS.  Such an emergence of new states can happen through different ways, two of the most important ones being:
the pitchfork bifurcation where a new  fixed-point (equilibrium) emerges from an already existing and the saddle-node bifurcation in which a stable-unstable pair of fixed-points
emerges spontaneously.

Two characteristic cases of such bifurcations are shown in Figs. \ref{larger1}(a),(e),(i) and \ref{larger1}(b),(f),(j) depicting the results for $N=4,~\nu=1$ and $N=4,~\nu=1/2$,
respectively. 
In the former case  (Figs. \ref{larger1}(a),(e),(i)) starting from the low-$f$ regime a new solution branch (green (gray) lines) 
appears at a critical value $f_A$, destabilizing the branch of low $f$ (red (black) lines) 
and becoming the new GS of the system for a very narrow $f_A<f<f_B$ interval. 
Beyond a certain value of $f=f_B$ the low-$f$ branch (red (black)) stabilizes again (probably through a collision with another state) 
and reestablishes itself as the GS of the system. These facts are reflected by the GS  energy difference $\Delta E$ (blue circles), characterizing the bending response of the system
in its respective GS configuration, causing its discontinuities at  $f=f_A,f_B$ (Fig. \ref{larger1} (e)). 

The case of $\nu=1/2$
(Figs. \ref{larger1}(b),(f),(j)) is more complex. The two saddle-node bifurcations occurring around $f\approx 2$ (Fig. \ref{larger1}(b)) give rise to pairs of stable-unstable branches and in turn to 
stable equilibrium states of an 
entirely different character than the low-$f$ state (red (black) line). Their energies cross the GS energy at certain $f$  values, $f=f_A$ and $f=f_B$ respectively,
resulting in them becoming the new GSs. Consequently, both the 
GS interparticle distances (blue circles) (Fig. \ref{larger1} (b)) and the GS $\Delta E$ (blue circles) (Fig. \ref{larger1} (f)) acquire a discontinuous form with small gaps at the points of the energy 
crossings $f=f_A,f_B$.

For larger values of $N$ the behaviour for large $f$ becomes more and more complex (two lowest rows of Fig. \ref{larger1}) with a plethora of saddle-node bifurcations leading
to multiple fixed-point branches (equilibrium states) whose energies cross the lowest energy at different values of $f$ ($f_A,~f_B,~f_C,~f_D$).
These crossings result in the GS consisting of parts of different equilibrium branches as a function of $f$ which gives the impression
of  discontinuous changes in the GS structure as $f$ increases (blue circles in Figs. \ref{larger1}(c),(d)), reflected also by  the GS $\Delta E$ 
 (blue circles in Figs. \ref{larger1}(g),(h)).

The above results suggest that the investigation of the GS bending response of the helically confined charges for large $f$ becomes a highly demanding and very case-specific task.
This is due to the massive dependence of $\Delta E$ on the GS configurations, resulting in it changing irregularly for increasing $f$, following the plethora
of energy crossings emerging through the increasing complexity of the potential landscape. 
Although it is difficult from the above results to provide general conclusions about the bending response 
of our system for large values of $f$, it seems that there is the tendency for $\Delta E$ to be overall negative at least for small particle numbers 
(Fig.~\ref{larger1} (e),(g),(h)) indicating that these systems are prone to bending with respect to $\phi_0 = 0$.
One of the most unifying features in the large-$f$ region is the character of the
profiles of $\Delta u_i$ (Figs. \ref{larger1} (i)-(l)). In all the cases, for sufficiently high $f>f_B,(f_D)$,
the $\Delta u_i$ profile becomes rather localized and non-uniform, developing
one (Figs. \ref{larger1} (i)-(k)) or several (Figs. \ref{larger1} (l)) peaks 
in contrast to its low-$f$ limit ($f<f_A$), where it is rather uniform, with all the $\Delta u_i$ acquiring similar values. From such a profile we can conclude that the particles tend to fragment into smaller chains consisting of
particles close to each other (small $\Delta u_i$) which are separated by large $\Delta u_i$ intervals similar to those observed for helically confined charges under a constant driving \cite{Zampetaki2017}. The extreme example
of such a case, for very large $f$, would be the number of particles splitting into two approximately equal 
parts with half the particles occupying the first winding and the other half the last winding of the helix.

Before closing this section, let us finally comment on the reaction of our system to different degrees of bending, beyond the small bending ($\bar{R}=50$)
addressed so far.

\begin{figure}[htbp]
\begin{center}
\includegraphics[width=1\columnwidth]{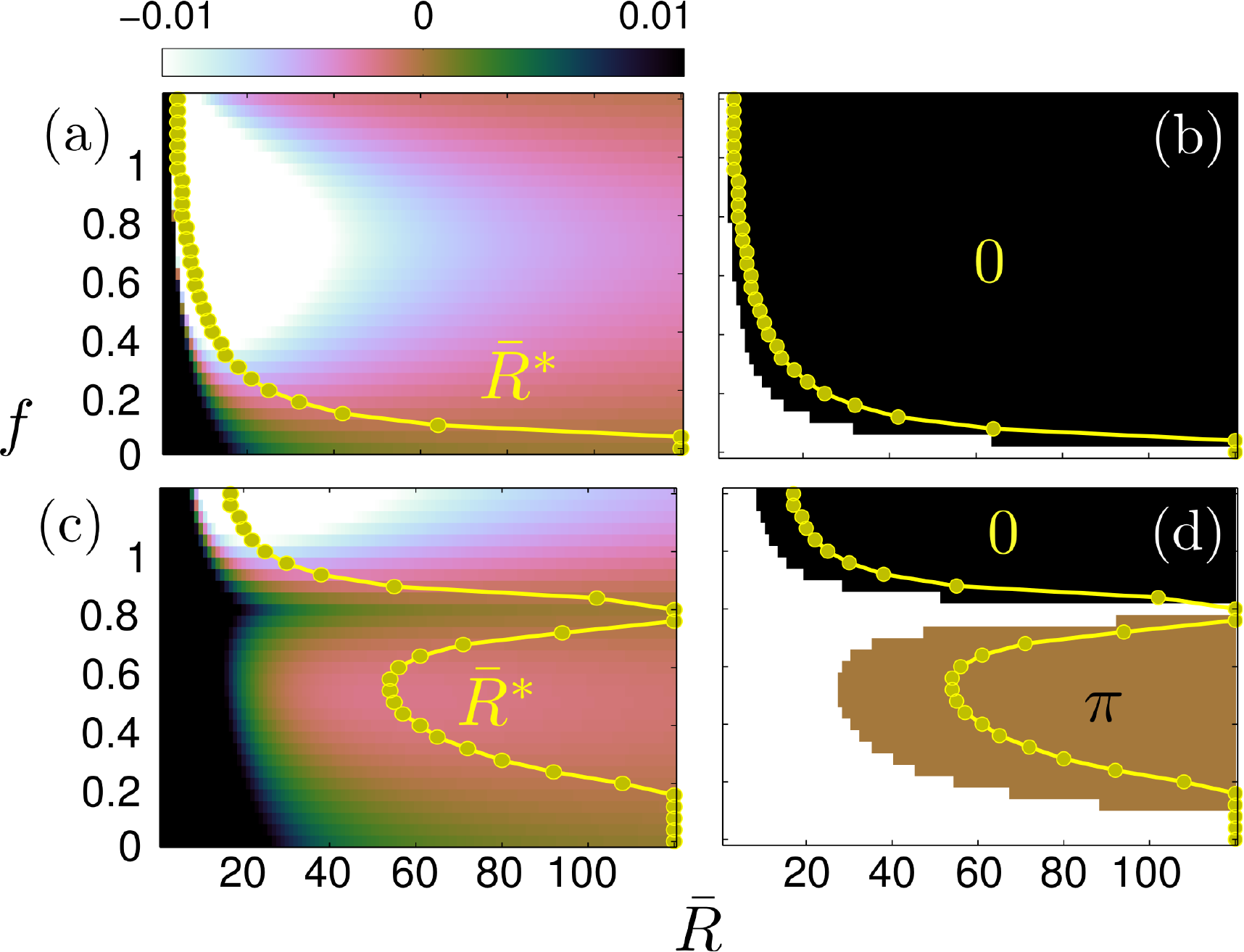}
\end{center}
\caption{\label{Ropt} (color online) 
 (a), (c) Dependence of the minimum  energy difference $\Delta E^*$ (depicted by color) between the bent and the unbent helical chain (resulting from a scan over the different values of $\phi_0 \in [0,2\pi)$)
on  $\bar{R}$  and  $f$ for (a) $N=10, ~\nu=1$, (c) $N=35, ~\nu=1$.
 (b),(d) Dependence of the preferable bending angle $\phi_0^*$ on 
$\bar{R}$ and $f$ for (b) $N=10, ~\nu=1$, (d) $N=35, ~\nu=1$, with the white regions corresponding to cases where bending is unfavorable.
The yellow line with the circles depicts the optimal value $\bar{R}^*$ of $\bar{R}$, giving the optimal degree of curvature,
as a function of $f$ for $N=10, \nu=1$ (subfigures (a),(b)) and $N=35, \nu=1$ (subfigures (c),(d)).}
\end{figure}

\begin{center}
{VII. \textbf{REGIME OF LARGER BENDING CURVATURE}}
\end{center}
We have seen in Sec. IV that there exist many cases for which the minimum energy difference $\Delta E^*$ is negative (see Fig. \ref{difends} (c)),
implying that a slight bending of the helical chain of charges is favorable. However, we have not examined so far up to which degree of bending this is still true, 
and what the energetically optimal curvature for our system is.
To provide an answer to these questions we have examined
the energy difference $\Delta E^*$ of the straight configuration from the optimally bent one ($\phi_0=\phi_0^*$) for $\nu=1$ and two different values of $N=10,35$
as a function of the geometrical parameter $f$ and the parameter $\bar{R}$ controlling the radius of curvature $R$ (Eq. \ref{Rcur}).
The results are shown in Fig.~\ref{Ropt}.

For the cases chosen  the regimes of $f$ values for which  $\Delta E^*$ is negative are different (Fig.~\ref{Ropt} (a),(c)).
This is expected from the oscillating pattern of $\Delta E^*$ in Fig.~\ref{difends}(c), where it is obvious that for different values of $N$
there is a different dependence of $\Delta E^*$ on $f$.
In both the cases of Fig. \ref{Ropt} (a) and (c) for very small $f$ we see that $\Delta E^*>0$ and bending is unfavorable.
For small $N$ (Fig. \ref{Ropt} (a)) we observe that with increasing $f$ the optimal value $\bar{R}^*$ of $\bar{R}$ decreases, approaching $\bar{R}^*=1$ for $f=1.2$.
This means that for large enough values of $f$ the energetically optimal structure tends to that
of a closed ring which corresponds to $\bar R^* =1$. Additionally, for every value of $f$ the preferable bending angle is towards the $+x$ direction ($\phi_0^*=0$) as depicted in Fig. \ref{Ropt} (b).

Although the aforementioned tendency with $\bar{R}^*$ overall decreasing with increasing $f$ is similar for larger $N$,
we observe for the case of $N=35$ (Fig.~\ref{Ropt} (c)) that this decrease can be non-monotonous. 
In this case $\bar{R}^*$, after 
 reaching a minimum value for $f\approx 0.5$, increases again with $f$ until $f\approx 0.8$ and from then on it continues to steadily decrease for even larger  values of $f$. This results 
 in the appearance of a peak in $\bar{R}^*$ as a function of $f$ which as shown in Fig. \ref{Ropt} (d) coincides with the change of the optimal bending angle from $\phi_0^*=\pi$
 to $\phi_0^*=0$.  Therefore in this case for increasing $f$ the system bends more and more in the $-x$ direction until a maximum bending is reached. From then on the optimal degree 
 of bending in the $-x$-direction decreases  until bending becomes totally unfavorable ($\Delta E^*>0$) from which point on the optimal bending again starts to increase but now in the $+x$ direction.
 Note that as the number of particles $N$ becomes larger, the optimal bending for the same value of $f$ decreases (compare Figs. \ref{Ropt} (a) and (c)), leading once more to the conclusion 
 that for large systems and small values of $f$ bending is overall unfavorable.

 \begin{center}
 \textbf{VIII. CONCLUSIONS AND OUTLOOK}  
\end{center}

We have shown that the bending response of a finite helical chain of charges displays a large variety depending on its geometrical properties, its filling and its size.
We have found that the energy difference $\Delta E$ between the bent and the straight conformation of the helix features unusual scaling with respect to the radius of curvature $R$ and the system length $L$, preventing the definition of quantities normally used to describe the bending response of the system, such as the bending rigidity and the persistence length. Therefore, we have chosen to examine the behavior
of $\Delta E$ in order to extract information about the bending response of the helical chain of charges.

Analytical considerations showed that $\Delta E$ consists of two dominant terms: an oscillatory one, weighted by the radius of the helix
and depending strongly on the GS configuration of the system, and a strictly positive one, weighted by the pitch and rather insensitive to the details of the GS.
The interplay between these contributions determines the magnitude and the sign of $\Delta E$.

For a radius  smaller or comparable to the pitch, the helical chain possesses a single equilibrium configuration, the GS of the system, which changes smoothly with the radius-to-pitch ratio 
$f$, the filling $\nu$ and the number of particles $N$. This yields a smooth variation of $\Delta E$ for varying parameters,
which displays particularly interesting features in the case of commensurate fillings.
In this case and assuming a bending in a fixed direction we have found that $\Delta E$ exhibits oscillations with a filling-dependent frequency as a 
function of the particle number. The origin of these oscillations has been shown to be the deviation of the GS configuration from the equidistant (uniform) one dictated by
the combination of the long-range interactions and the fixed boundary conditions. In a more physically sound perspective such oscillations account for oscillations in 
the bending direction of the helical chain, with $\Delta E$ being overall negative.  The bending direction is determined by the side of the helix with the highest 
charge occupancy since, as similarly discussed for the case of charged cylinders \cite{Nguyen1999}, it is the imbalance between the number of charges on the outer and the inner side which
can lead to a bending conformation being favored.

For values of the radius significantly higher than the values of the pitch, the potential landscape of our system becomes increasingly complex causing the emergence of 
multiple stable equilibria which for certain parameters can become the GS of the system. Usually these possess very different structures yielding a discontinuous behaviour
of the GS properties with the radius-to-pitch ratio $f$. This discontinuity is reflected in the variation of $\Delta E$ with $f$ and results in 
its very case-specific character.

In summary, we have shown in this work how the complexity of the energy landscape, the non-uniformity of the charge distribution and the finite size can affect the bending response 
of a charged helix. Throughout this study we have addressed only the electrostatic contribution to $\Delta E$.  For a complete analysis also the elastic 
bending response of the helical confining filament should be taken into account. This could be achieved by employing the standard techniques \cite{Antman2005,Balaeff2006, Chen2011},
given particular information about the material of the helical confining segment in the context
of  a specific experimental realization. Possible experimental setups for which our model could apply 
include, except for  charged helical macromolecules \cite{Kornyshev2007}, also charges confined in free-standing nanostructures \cite{Prinz2000, Zhang2009,Lee2014, Ren2014} or even charged beads confined on helical wires as in 
mechanical setups used in the study of polymers \cite{Reches2009,Tricard2012}.

Returning to the theory side, further studies could investigate the bending response of the helical chain of charges in the presence 
of a dielectric medium or an electrolyte solution, relevant 
for biological considerations  \cite{Kornyshev2007}. They could additionally focus on the effects of driving of the spatial conformation of the helix, provided e.g. by a time-dependent pitch.
As a direct extension of the current work, also the response of our system to different elastic deformations \cite{Golestanian1999} would be of interest.

\begin{center}
{ \textbf{APPENDIX: ASYMPTOTIC BEHAVIOUR OF $\Delta E^{(3)}, ~\Delta E^{(4)}$ WITH INCREASING SYSTEM SIZE}}
\end{center}
As mentioned in the main text (Sec. III) the two major contributions $\Delta E^{(3)}, ~\Delta E^{(4)}$ to the energy difference $\Delta E$ scale differently with the particle number $N$.
We provide here an analytical estimation of this dependence assuming an approximately  equidistant GS configuration $\{u_{i}^{(0)}=u_{i}^ {(eq)}=2\pi(i-1)/\nu\}$, with $\nu=(N-1)/M$ the
filling. Using this assumption as well as the trigonometric identity for the sum of cosines and  changing the summing indices $1 \leq i,j \leq N$ to 
$1 \leq l=\abs{i-j} \leq N-1$, $1 \leq j \leq N-l$, 
eqs. (\ref{de3}),(\ref{de4}) take the form

\begin{equation}
\Delta E^{(3)}= -\frac{b_1}{N-1}\sum_{l=1}^{N-1}\sum_{j=1}^{N-l}\frac{l^2\cos \left[(l+2j-2)\frac{\pi}{\nu}\right]\cos \frac{l\pi}{\nu}}{\left[2r^2\left(1-\cos\frac{l\pi}{\nu}\right)+\frac{h^2 l^2}{\nu^2}\right]^{3/2}} \label{de3a}
\end{equation}

\begin{equation}
\Delta E^{(4)}= \frac{b_2}{(N-1)^2}\sum_{l=1}^{N-1}\frac{(N-l)l^4}{\left[2r^2\left(1-\cos\frac{l\pi}{\nu}\right)+\frac{h^2 l^2}{\nu^2}\right]^{3/2}}. \label{de4a}
\end{equation}
In order to make the $N$-dependence explicit in the above equations we have used Eq. (\ref{Rcur}) with $\Phi=2\pi(N-1)/\nu$, resulting in 
$b_1=\frac{2 \lambda r \pi h}{\bar{R} \nu}$ and $b_2=\frac{\lambda h^2 \pi^2}{6 \bar{R}^2 \nu^2}$.

Regarding the magnitudes $\abs{\Delta E^{(3)}},~\abs{\Delta E^{(4)}}$ these are bounded from above as follows
\begin{eqnarray}
\abs{\Delta E^{(3)}}&\leq& \frac{c_1}{N-1}\sum_{l=1}^{N-1}\frac{N-l}{l}\nonumber \\ 
&\approx& c_1\frac{N\log(N-1)+(\gamma-1)N+1}{N-1} \label{de3a2}
\end{eqnarray}

\begin{equation}
\abs{\Delta E^{(4)}}\leq \frac{c_2}{(N-1)^2}\sum_{l=1}^{N-1} (N-l)l  
= c_2\frac{(N+1)N}{6(N-1)}\label{de4a2}
\end{equation}
with $\gamma\approx 0.5772$ the Euler-Mascheroni constant, $c_1=\frac{2 \lambda r \pi  \nu^2}{\bar{R} h^2}$, $c_2=\frac{\lambda h^2 \pi^2 \nu}{6 \bar{R}^2 h}$.
Here it was employed that $\abs{\cos \frac{l\pi}{\nu}} \leq 1$, where the equality holds for the special cases of commensurate fillings $\nu=1/n$ or 
in the limit of very dense systems $\nu \rightarrow \infty$.
Note that for the calculation of Eq.~(\ref{de3a2}) we have used  the approximation 
$\sum_{l=1}^{N-1}\frac{1}{l} \approx \log (N-1)+\gamma$ \cite{Abramowitz1970}.

The bounds (\ref{de3a2}),(\ref{de4a2}) yield 
\begin{equation}
\abs{\Delta E^{(3)}}=\mathcal{O}(\log N),~~ \abs{\Delta E^{(4)}}=\mathcal{O}(N) \label{bou1}
\end{equation}
meaning that $\abs{\Delta E^{(4)}}$ increases in general significantly faster with the number of particles $N$ than $\abs{\Delta E^{(3)}}$.

Going back to eqs. (\ref{de3a},\ref{de4a}), it can be seen that except for the denominator that is common for $\Delta E^{(3)}$ and $\Delta E^{(4)}$, 
the term $\Delta E^{(4)}$ does not depend in any other way on $r$ and $\cos \frac{l\pi}{\nu}$ (Eq. (\ref{de4a})).
Thus, the extracted bound for $\Delta E^{(4)}$ (Eq. (\ref{de4a2})) is much tighter than the one for $\Delta E^{(3)}$ (Eq. (\ref{de3a2}))
and it turns out that it approximates in general very well the actual value of $\Delta E^{(4)}$ (Eq. (\ref{de4a})). As a consequence we have
that $\Delta E^{(4)}$ is almost $r$-independent and that approximately $\Delta E^{(4)} \propto \frac{\nu}{\bar{R}^2} N$ (Eq. (\ref{de4a2}))
 for any value of the filling $\nu$. 
In contrast, $\Delta E^{(3)}$ depends strongly on the values of $\cos\frac{l\pi}{\nu}$. These values are in general (apart from the cases of commensurate or very high fillings where $\cos \frac{l\pi}{\nu}=1$)
scattered in $[-1,1]$, leading to the sum in Eq. (\ref{de3a}) approximately averaging to zero.
Thus $\Delta E^{(4)} \propto  \frac{\nu}{\bar{R}^2} N$ for large $N$, increasing linearly both with the size $N$ and with the filling $\nu$,
whereas the contribution of $\Delta E^{(3)}$ 
is generally much smaller, unless in the commensurate or densely-filled cases where $\Delta E^{(3)} \propto \frac{\nu^2}{\bar{R}}\log N$.

In the densely-filled limit, especially as $\nu \rightarrow \infty$, $\Delta E^{(3)}$ is expected to 
be the dominant contribution to $\Delta E$
due to its faster scaling with the filling ($\nu^2$ vs. $\nu$) and its slower decay with $\bar{R}$ ($1/\bar{R}$ vs. $1/\bar{R}^2$), causing $\Delta E \leq 0$.
For commensurate fillings, although $\Delta E^{(3)}$ is dominant for small $N$
(since it decays more slowly with $\bar{R}$), as $N$ becomes larger  the term $\Delta E^{(4)}$ overtakes it 
for constant $\bar{R}$ due to its faster scaling with the particle number ($N$ vs. $\log N$). Thus even in the case of commensurate fillings where
$\Delta E^{(3)}$ attains its most negative value (Eq. (\ref{de3a})), the strictly positive $\Delta E^{(4)}$ (Eq. (\ref{de4})) will prevail
for large particle numbers $N\gg 1$, yielding overall a positive $\Delta E$.
All these features have been indeed observed in our numerical results in Sec. IV (Fig.~\ref{be_ri123},~\ref{ofil2}). 

To conclude, the term $\Delta E^{(3)}$ causing a different bending response of a helical chain of ions from that of a linear chain 
becomes substantial only for small particle numbers $N$ (Eq. (\ref{bou1})), commensurate fillings $\nu=1/n$ and/or very dense systems $\nu \rightarrow \infty$.
Note, however, that all the arguments of this section are based on the assumption of an approximately equidistant GS configuration, valid only for 
small values of $f=\frac{2\pi r}{h}$. For large values of $f$ (large helix radius $r$) this assumption is strongly violated 
resulting in an \textit{a priori} unpredictable behaviour of the terms $\Delta E^{(3)},~\Delta E^{(4)}$.

\begin{center}
{ \textbf{ACKNOWLEDGEMENTS}}
\end{center}
A. Z. thanks C. V. Morfonios for useful discussions.

\end{document}